\documentclass[journal,twoside,web]{ieeecolor}
\input{3-compilation/packages}
\theoremstyle{plain}
\newtheorem{assumption}{Assumption}
\newtheorem{theorem}{Theorem}
\newtheorem{lemma}{Lemma}
\newtheorem{proposition}{Proposition}
\newtheorem{corollary}{Corollary}
\theoremstyle{definition}

\renewcommand{\qedsymbol}{\ensuremath{\square}}

\newcommand{\bR}{\mathbb{R}}
\newcommand{\bN}{\mathbb{N}}
\newcommand{\one}{\mathbf{1}}
\newcommand{\cX}{\mathcal{X}}
\newcommand{\cU}{\mathcal{U}}
\newcommand{\cD}{\mathcal{D}}
\newcommand{\cC}{\mathcal{C}}
\newcommand{\cB}{\mathscr{B}}
\newcommand{\cS}{\mathcal{S}}
\newcommand{\cSB}{\cS_{\cB}}
\newcommand{\I}{\mathcal{I}}
\newcommand{\J}{\mathcal{J}}
\newcommand{\B}{\mathcal{B}}
\newcommand{\tr}{^{\top}}
\newcommand{\norm}[1]{\lVert #1\rVert}
\newcommand{\sigmin}{\sigma_{\min}}
\newcommand{\eps}{\varepsilon}
\DeclareMathOperator*{\argmin}{arg\,min}
\DeclareMathOperator{\Lip}{Lip}
\DeclareMathOperator{\diag}{diag}
\DeclareMathOperator{\col}{col}
\DeclareMathOperator{\range}{range}
\DeclareMathOperator{\rank}{rank}
\DeclareMathOperator{\Reg}{Reg}
\DeclareMathOperator{\diam}{diam}

\newcommand{\fd}{f_{\mathrm{d}}}
\newcommand{\ug}{\underline{g}}
\newcommand{\piid}[1]{\pi_{\mathrm{nom},#1}}
\newcommand{\decay}{\theta_{\Lambda}}
\newcommand{\sn}[1]{s_{\mathrm{n},#1}}
\newcommand{\sno}{\bar{s}_{\mathrm{n}}}
\newcommand{\Lsf}{L_{\mathrm{sf}}}

\begin{document}
\title{Robust Safe Control Under Uncertainty: Estimation, Contraction, and Regret Bounds}
\author{Changrui Liu, \IEEEmembership{Graduate Student Member, IEEE}, Liang Wu, Anil Alan, \\ Shengling Shi, \IEEEmembership{Member, IEEE}, and Bart De Schutter, \IEEEmembership{Fellow, IEEE}
\thanks{Corresponding authors: Changrui Liu and Shengling Shi.}
\thanks{Changrui Liu, Anil Alan, Shengling Shi, and Bart De Schutter are with the Delft Center for Systems and Control, TU Delft, Delft, The Netherlands. (e-mail: \{c.liu-14, a.alan, shengling.shi, b.deschutter\}@tudelft.nl). }
}

\maketitle

\begin{abstract}
Disturbance uncertainty requires safety filters based on control barrier functions (CBFs) to leave a robustness margin, which can alter the nominal control input and degrade closed-loop performance. This paper studies safety and the resulting performance loss for discrete-time nonlinear systems with additive disturbances and polytopic input constraints. We develop a causal disturbance observer and construct two robust barrier conditions. The first uses a time-dependent bound on the estimation error. The second uses a constant margin when the disturbance increments are uniformly bounded and moves the requirement on the initial error from the online filter to an explicit initialization condition. For affine barriers, both filters are quadratic programs. We bound their input deviations from an oracle filter with access to the current disturbance and derive sufficient conditions for uniform contraction of the oracle closed loop. The contraction analysis accounts for state-dependent actuation, unmatched disturbances, and active physical input limits. Combining the sensitivity and contraction estimates gives finite-time dynamic regret bounds that quantify the effects of observer convergence, disturbance variation, and robustness margins. With sublinear cumulative certified disturbance variation and observer innovation error, the bounds yield sublinear regret.
\end{abstract}


\section{Introduction}
\label{sec:introduction}
\IEEEPARstart{S}{afety}-critical control aims to keep the state of a dynamical system within a prescribed set. Hamilton--Jacobi reachability analysis and control barrier functions (CBFs) provide two approaches to this objective~\cite{mitchell2005reachable,ames2019control}. CBFs express forward invariance through conditions on the control input~\cite{ames2016control}. A safety filter can then enforce these conditions by modifying a nominal command only when it violates a constraint. This architecture separates the design of a performance-oriented controller from the enforcement of safety. It can be combined with classical feedback, model predictive control (MPC)~\cite{zeng2021safety}, or learned policies obtained through approximate MPC and reinforcement learning~\cite{liu2026approximate,cohen2023safe}.

Most CBF designs have been developed for continuous-time systems~\cite{ames2016control,lopez2020robust,wang2025safe}. In applications such as robotic locomotion, energy management, and autonomous driving, controllers often use discrete-time models and act at sampling instants~\cite{agrawal2017discrete,liu2026approximate,katriniok2023discrete}. Discrete-time CBFs impose conditions on the successor state~\cite{agrawal2017discrete,cosner2023robust,cosner2024generative}. The controller must therefore account for the disturbance acting over the next transition using information available at the current step. This timing also matches that of a disturbance observer, which can update its estimate after each measured transition.

Unknown disturbances motivate robust safety filters. A margin based on the full disturbance set can ensure safety, but it may require unnecessary changes to the nominal input when recent measurements are informative about the current disturbance. Robust adaptive approaches use parameter estimates or disturbance observers to reduce the uncertainty the filter must accommodate~\cite{lopez2020robust,alan2022disturbance}. In discrete time, stochastic CBF methods give safety guarantees under probabilistic disturbance models~\cite{cosner2023robust}, while robust adaptive CBFs address parametric model uncertainty and disturbances~\cite{liu2027robust}. Here we consider an unknown, nonparametric disturbance sequence with a deterministic bound on its increments. When the actuation matrix has fewer columns than rows, an input correction can cancel only the component in its range. The remaining component must be accounted for in the barrier condition, and the compensation must respect the physical actuator limits.

Robustification can change both the input selected by a safety filter and the subsequent state trajectory. Regret and competitive-ratio analyses quantify related performance losses in predictive control under model mismatch or imperfect inputs~\cite{liu2026certainty,liu2025regret}. Regret has also been studied in online control with disturbances~\cite{agarwal2019online}, safe control with hard constraints~\cite{martin2022safe}, and learning with stochastic CBFs~\cite{luo2022safe}. These results motivate a comparison that holds the nominal controller and the barrier requirements fixed while changing the disturbance information available to the filter. We use as a comparator an oracle filter with access to the current disturbance at each step. The resulting regret measures the cumulative cost difference attributable to operating the safety filter with estimated disturbance information and robustness margins.

We address two questions. Can the online barrier condition be designed without a bound on the initial estimation error, and what initialization does this require? How does the estimation error affect closed-loop regret, and how can the conditions needed for the regret analysis be checked? The contributions are as follows.
\begin{enumerate}
\item We derive two deterministic safety conditions for Lipschitz continuous barrier functions using a discrete-time disturbance observer. One uses a time-dependent estimation-error bound. The other uses a margin determined by the disturbance increments and the observer gain, with an explicit condition on the initial estimation error.
\item For affine barriers and polytopic input constraints, we quantify the sensitivity of the safety-filter quadratic program (QP) to the disturbance estimate and the robustness margin. We derive the Jacobian of the oracle closed loop across active sets and obtain contraction tests that use known model and constraint data. The analysis builds on the piecewise-smooth viewpoint used for continuous-time barrier-certified systems~\cite{marvi2024robust}.
\item We combine the sensitivity and contraction estimates to bound the finite-time dynamic regret of both robust filters. The bounds distinguish the transient from initialization, the cumulative disturbance variation and innovation error, and the amplification due to the oracle contraction factor.
\end{enumerate}

Section~\ref{sec:preliminaries} introduces the model, the barrier constraints, and the disturbance observer. Section~\ref{sec:problem_formulation} defines the oracle comparator and the regret. Section~\ref{sec:safe_control} develops the robust filters, and Section~\ref{sec:theoretical_analysis} presents the sensitivity, contraction, and regret analyses. Section~\ref{sec:conclusion} concludes the paper.

\section{Preliminaries}
\label{sec:preliminaries}
\subsection{Notation}
The nonnegative integers are denoted by $\bN$, and $[k]=\{1,\ldots,k\}$ for a positive integer $k$. The symbols $I_k$ and $\one_k$ denote the identity matrix and the all-ones vector of dimension $k$. A subscript consisting of an index set selects rows of a matrix or entries of a vector. We use $\col$ for vertical stacking and $\diag$ for the diagonal matrix with the specified entries. Vector inequalities are componentwise. For symmetric matrices, $\succeq$ and $\succ$ denote positive semidefinite and positive definite ordering. The Euclidean vector norm and its induced matrix norm are both denoted by $\norm{\cdot}$. For a rectangular matrix $A$, $\sigmin(A)$ denotes the smallest of its $\min\{\text{rows},\text{columns}\}$ singular values, and $A^\dagger$ denotes its Moore--Penrose inverse.

For a differentiable matrix-valued map $A$, $DA(x)[v]$ is its directional derivative in the direction $v$. For a vector-valued map we also write $DF(x)v$. We use $\norm{DA(x)}=\sup_{\norm v=1}\norm{DA(x)[v]}$. When the time and disturbance are suppressed, state derivatives hold both fixed. For a map $F$ on a set $S$, $\Lip_x(F;S)$ denotes its least Lipschitz constant with respect to the state on $S$. We use $\Gamma_T(a)=\sum_{k=0}^{T-1}a^k$ for the finite geometric sum, including $\Gamma_0(a)=0$. Empty row stacks, products with zero inner dimension, and sums over empty index sets have their usual zero-dimensional or zero values.

\subsection{System Settings \& Nominal Controller}
\label{subsec:settings}
Consider the input-affine discrete-time system
\begin{equation}
    \label{eq:sec2.b-model}
    x_{t+1}=\underbrace{\fd(x_t)+g(x_t)u_t}_{f(x_t,u_t)}+d_t,
\end{equation}
where $x_t\in\cX\subseteq\bR^n$ and $u_t\in\cU\subseteq\bR^m$ are the state and the applied input at time $t$, respectively. The sets $\cX$ and $\cU$ are compact, and $\cU$ is a polytope containing the origin in its interior:
\begin{equation}
    \label{eq:input_set_polytope}
    \cU=\{u\in\bR^m:E_u u\leq\one_{c_u}\}.
\end{equation}
Here $E_u\in\bR^{c_u\times m}$ and $c_u$ is the number of input inequalities. This representation includes box constraints, which are often used to model actuator limits~\cite{ames2016control,lopez2020robust}. The drift $\fd:\bR^n\to\bR^n$ and input coupling $g:\bR^n\to\bR^{n\times m}$ define the nominal dynamics $f$. The term $d_t\in\cD\subseteq\bR^n$ is an unknown, nonparametric additive disturbance in a known nonempty compact convex set $\cD$. Model~\eqref{eq:sec2.b-model} represents known dynamics plus plant--model mismatch, as in cruise control, robotics, and energy systems~\cite{taylor2020adaptive,agrawal2017discrete,liu2026approximate,wang2025safe}. We consider exogenous sequences $\{d_t\}_{t\geq0}$ so that two trajectories can be compared under the same disturbances.

The system state is required to remain in the safe set $\cSB$ characterized by a collection of affine barrier functions $\cB=\{B_1,\ldots,B_{N_B}\}$, with
\begin{equation}
    \label{eq:barrier_affine_closed_form}
    B_i(x)=b_i\tr x-s_i,\qquad b_i\in\bR^n\setminus\{0\},\quad s_i\in\bR.
\end{equation}
By the Cauchy--Schwarz inequality, $B_i$ is Lipschitz continuous with constant $L_{B,i}=\norm{b_i}$. Each barrier defines $\cS_i=\{x\in\bR^n:B_i(x)\geq0\}$, and the joint safe set is $\cSB=\bigcap_{i=1}^{N_B}\cS_i\subseteq\cX$. We write $\cC=\cSB$ and assume that it is a nonempty compact polytope with nonempty interior. Its defining rows include any state bounds needed for $\cC\subseteq\cX$. Affine barriers can describe interval constraints and polyhedral inner approximations of convex safe sets. Examples include bounds on stored energy and machine speed~\cite{liu2026approximate,liu2027robust}.

Stack the row normals as $H=\col(b_i\tr)_{i\in[N_B]}$ and the offsets as $s=\col(s_i)_{i\in[N_B]}$. We use retention factors $\alpha_i\in(0,1]$ and define
\begin{equation}
 A_\alpha=\diag(\alpha_i)_{i\in[N_B]},\qquad
 \tau(x)=A_\alpha Hx+(I_{N_B}-A_\alpha)s.
 \label{eq:alpha}
\end{equation}
The discrete-time barrier condition $B_i(x_{t+1})\geq\alpha_i B_i(x_t)$ for every $i$ is equivalent to $Hx_{t+1}\geq\tau(x_t)$. This condition gives forward invariance of $\cC$ when enforced at every step. Section~\ref{sec:safe_control} first derives the robust conditions for Lipschitz continuous barriers and then specializes them to~\eqref{eq:barrier_affine_closed_form}.

At each time $t$, given the measured state $x_t$ and the applied input $u_t$, the system incurs a stage cost $\ell_t(x_t,u_t)$, where $\ell_t:\cX\times\cU\to\bR$. We assume that the full state is measured without noise.
\begin{assumption}[Lipschitz Continuous Stage Cost]
\label{ass:cost_lipschitz}
There are $L_{\ell,x},L_{\ell,u}>0$ such that, for all $t\in\bN$, $x',x''\in\cX$, and $u',u''\in\cU$,
\begin{multline*}
 |\ell_t(x',u')-\ell_t(x'',u'')|\leq
 L_{\ell,x}\norm{x'-x''}+L_{\ell,u}\norm{u'-u''}.
\end{multline*}
\end{assumption}
This assumption is standard in optimal control~\cite{lin2021perturbation,liu2025regret}. It includes linear costs~\cite{rawlings2012fundamentals}, $1$- and $\infty$-norm costs~\cite{borrelli2017predictive}, and quadratic costs on compact domains with uniformly bounded weights~\cite{rawlings2017model,liu2024stability}. We are given a continuous, possibly time-varying nominal controller $\piid{t}:\bR^n\to\bR^m$ designed using $f$ to achieve the stage-cost objective. It may return commands that violate the safety or physical input constraints. The safety filter uses $\piid{t}(x_t)$ as its reference command.

\begin{assumption}[Domain and actuation]
\label{ass:domain}
The functions $\fd$, $g$, and $\piid{t}$ for every $t\in\bN$ are $C^1$ on a neighborhood of $\cC$. There is a known $\ug>0$ such that $g(x)\tr g(x)\succeq\ug^2 I_m$ for every $x\in\cC$.
\end{assumption}

Under Assumption~\ref{ass:domain}, $m\leq n$. Define the left inverse and the projector onto the unmatched directions by
\begin{equation}
 g^\dagger=(g\tr g)^{-1}g\tr,\qquad P=I_n-gg^\dagger .
 \label{eq:projectors}
\end{equation}
Then $g^\dagger g=I_m$, $\norm{g^\dagger}\leq1/\ug$, and both $P$ and $gg^\dagger$ are orthogonal projectors with norms at most one. In particular,
\begin{equation}
 d=\underbrace{gg^\dagger d}_{\text{matched}}+\underbrace{Pd}_{\text{unmatched}},
 \qquad g\tr Pd=0.
 \label{eq:decomposition}
\end{equation}
Given a virtual command $u$ and a disturbance value $a$, the applied command is $w=u-g^\dagger(x)a$. The successor predicted with $a$ is $\fd(x)+g(x)u+P(x)a$. Exact compensation therefore removes the matched disturbance from this expression. The input constraint in the virtual variable is $E_u u\leq\one_{c_u}+E_u g^\dagger(x)a$. For square actuation $P=0$, but this constraint can still make the optimizer and the state update depend on $a$ when a physical limit is active. Even with no physical limit active, the applied input retains the compensation term.

For the affine barriers, define the common QP data and the virtual optimizer
\begin{equation}
 \begin{gathered}
 M(x)=\begin{bmatrix}-Hg(x)\\ E_u\end{bmatrix},\\
 \beta(x,a)=\begin{bmatrix}
 H(\fd(x)+P(x)a)-\tau(x)\\
 \one_{c_u}+E_u g^\dagger(x)a
 \end{bmatrix},
 \end{gathered}
 \label{eq:data}
\end{equation}
\begin{equation}
 \psi_t(x,a,r)=\argmin_{M(x)u\leq\beta(x,a)-r}\tfrac12\norm{u-\piid{t}(x)}^2 ,
 \label{eq:qp}
\end{equation}
where $r\in\bR^{N_B+c_u}$ has zero entries in the physical input block. The oracle is $\pi_{\mathrm{o},t}(x,d)=\psi_t(x,d,0)$. Its applied input $w_{\mathrm{o},t}$ and state update $F_t$ are
\begin{equation}
 \begin{aligned}
 w_{\mathrm{o},t}(x,d)&=\pi_{\mathrm{o},t}(x,d)-g^\dagger(x)d,\\
 F_t(x,d)&=\fd(x)+g(x)\pi_{\mathrm{o},t}(x,d)+P(x)d.
 \end{aligned}
 \label{eq:F}
\end{equation}
The equivalent QP in the physical variable $w$ is
\begin{equation}
 \begin{aligned}
 w_{\mathrm{o},t}(x,d)&=\argmin_{w}\tfrac12\norm{w-\big(\piid{t}(x)-g^\dagger(x)d\big)}^2\\
 \text{s.t.}\quad -Hg(x)w&\leq H(\fd(x)+d)-\tau(x),\\
 E_u w&\leq\one_{c_u}.
 \end{aligned}
 \label{eq:physicalqp}
\end{equation}
Feasibility gives $HF_t(x,d)\geq\tau(x)$ and $w_{\mathrm{o},t}(x,d)\in\cU$. The observer-based filters use the same QP with $a=\hat d_t$ and a tightened barrier block.

The following contraction results state when the oracle in~\eqref{eq:F} forgets initial-state differences uniformly. A contraction factor $q$ satisfies $\norm{F_t(x,d)-F_t(y,d)}\leq q\norm{x-y}$ for every $x,y\in\cC$, $d\in\cD$, and $t\in\bN$, with $q<1$. Section~\ref{subsec:contraction_analysis} specifies the constants and the active-row classes used in the tests. The detailed derivation is given in Appendices~\ref{app:geometry} and~\ref{app:certificate}.

\begin{theorem}[Uniform contraction of the oracle closed loop]
\label{thm:certificate}
Let Assumptions~\ref{ass:domain}, \ref{ass:feasible}, \ref{ass:licq}, and \ref{ass:bounds} hold. Let $C_*$ bound the oracle Jacobian after imposing the active physical input equalities, before imposing the active barrier equalities. Let $d_*$ bound the normal block induced by the active barriers. For branches with $0<r<n$ active barriers, obtain $p_*,\chi_*,\eps_*$ from~\eqref{eq:globalgeometry} using a certified geometry margin~\eqref{eq:eta} or its structural alternative. A sufficient contraction test is
\begin{equation}
 q_* = \sqrt{(p_*C_*+\chi_*d_*+\eps_*)^2+d_*^2}<1 .
 \label{eq:finaltest}
\end{equation}
If partial barrier activity is excluded by the QP data, use $q_* = \max\{C_*,d_*\}$. If every active pattern has $r=n$, use $q_*=d_*$, and if no barrier can become active, use $q_*=C_*$. Each test is uniform in $t$ and $d\in\cD$.
\end{theorem}

In the square case, define $W=gg\tr$ and suppose $\underline w I_n\preceq W(x)\preceq\bar w I_n$, with $0<\underline w\leq\bar w$ and common retention $\alpha_i=\alpha$. Let $q_U$ be the bound in~\eqref{eq:qU} for branches with active physical input limits. Let $c$ denote the nominal closed-loop Jacobian bound in Assumption~\ref{ass:bounds}.
\begin{corollary}[Square actuation with physical input limits]
\label{cor:conditioned}
With $p_b,\chi_b,\eps_b$ in~\eqref{eq:conditionedconstants}, the oracle is uniformly contractive if
\begin{equation}
 \max\{q_b,q_U\}<1,\qquad
 q_b=\sqrt{(p_bc+\chi_b\alpha+\eps_b)^2+\alpha^2}.
 \label{eq:corfull}
\end{equation}
If partial barrier activity is excluded on branches without active physical input limits, take $q_b=\max\{c,\alpha\}$ and omit the term for any class proved absent.
\end{corollary}
\begin{corollary}[Isotropic square actuation]
\label{cor:isotropic}
Under the conditions of Corollary~\ref{cor:conditioned}, if $gg\tr=a(x)I_n$ with $a(x)>0$, the bound on branches without active physical input limits is $q_b=\sqrt{c^2+\alpha^2}$. The same class reduction and the same $q_U$ apply.
\end{corollary}

\subsection{Discrete-Time Disturbance Observer}
\label{subsec:observer}
The one-step barrier condition depends on the disturbance itself. We use a first-order observer motivated by continuous-time disturbance estimation~\cite{das2016control} and driven by the input actually applied to~\eqref{eq:sec2.b-model}. Let $\xi_t\in\bR^n$ be its internal state. The update is
\begin{subequations}
\label{eq:observer}
\begin{align}
 \hat d_t&=\Lambda x_t-\xi_t,\\
 \xi_{t+1}&=\xi_t+\Lambda(f(x_t,u_t)-x_t+\hat d_t+\sn{t}),
\end{align}
\end{subequations}
where $\Lambda$ is symmetric and $0\prec\Lambda\prec I_n$. The term $\sn{t}$ represents a perturbation in the observer computation. The commanded update uses $f(x_t,u_t)-x_t+\hat d_t$, and the state measurement is exact. The estimate $\hat d_t$ is available when choosing $u_t$, while the new state $x_{t+1}$ enters the estimate at the following step. With exact innovation the observer uses the residual $x_{t+1}-f(x_t,u_t)=d_t$ only after that transition.

\begin{assumption}[Disturbance variation and observer error]
\label{ass:variation}
Known constants $\bar\Delta_d,\sno\geq0$ satisfy $\norm{d_{t+1}-d_t}\leq\bar\Delta_d$ and $\norm{\sn{t}}\leq\sno$ for all $t\in\bN$.
\end{assumption}
Let $e_t=d_t-\hat d_t$, $\Delta_{d,t}=d_{t+1}-d_t$, and $\omega_t$ be a known upper bound on $\norm{\Delta_{d,t}}+\norm{\Lambda\sn{t}}$, available by step $t$. The constant $\bar\omega=\bar\Delta_d+\lambda_{\max}(\Lambda)\sno$ always works. When sharper bounds are available, we take $0\leq\omega_t\leq\bar\omega$. Define
\begin{equation}
 A_\Lambda=I_n-\Lambda,\qquad
 \decay=\norm{A_\Lambda}=1-\lambda_{\min}(\Lambda).
 \label{eq:decay}
\end{equation}
The observer error obeys
\begin{equation}
 e_{t+1}=A_\Lambda e_t+\Delta_{d,t}+\Lambda\sn{t}.
 \label{eq:recursive_error}
\end{equation}
Consequently, if $E_0\geq\norm{e_0}$ is known, the recursion $E_{t+1}=\decay E_t+\omega_t$ supplies the bound
\begin{equation}
 \norm{e_t}\leq E_t=\decay^t E_0+
 \sum_{k=0}^{t-1}\decay^{t-1-k}\omega_k .
 \label{eq:estimation_error_bound}
\end{equation}
For $\omega_t=\bar\omega$ this is $E_t=(E_0-E_\infty)\decay^t+E_\infty$, where $E_\infty=\bar\omega/(1-\decay)$. The bound decays from $E_0$ when $E_0>E_\infty$ and increases to $E_\infty$ when $E_0<E_\infty$. An available choice for $E_0$ is $\max_{d\in\cD}\norm{d-\hat d_0}$. Appendix~\ref{app:observer} derives~\eqref{eq:recursive_error}--\eqref{eq:estimation_error_bound}. For a symmetric gain with spectrum in $(0,2)$ the same results use $\decay=\max_i|1-\lambda_i(\Lambda)|$.

\section{Problem Formulation}
\label{sec:problem_formulation}
Given $x_0\in\cC$ and the nominal command $\piid{t}(x_t)$, the online design must keep $x_t\in\cC$ and $u_t\in\cU$ for every time step. It may use the current state and the disturbance estimate from~\eqref{eq:observer}. Its benchmark has access to $d_t$ before solving~\eqref{eq:qp}. Both filters use the same reference command, barrier factors, and physical constraints. The oracle knows the current disturbance only.

\begin{assumption}[Filter feasibility]
\label{ass:feasible}
The oracle QP is feasible for every $(x,d)\in\cC\times\cD$. The robust QP being applied is feasible at each reachable state and observer estimate for the admissible initial conditions and disturbance sequences under consideration.
\end{assumption}
The robust feasibility requirement is for the chosen margin and, for the robustified barrier design, the initialization in Section~\ref{subsec:robustified}. It can be verified on a larger set of state--estimate pairs if one is available. Each feasible QP has a unique minimizer because the objective is strictly convex and the feasible inputs lie in the compact translate $\cU+g^\dagger(x)a$.

For the oracle define the set of active inequalities at $(x,d)$ by
\begin{equation}
 \mathcal A_t(x,d)=\{i:M_i(x)\pi_{\mathrm{o},t}(x,d)=\beta_i(x,d)\}.
 \label{eq:actualactive}
\end{equation}
In the time-varying case, we also use the same QP with a fixed reference command $z$ in place of $\piid{t}(x)$. Let
\begin{equation}
 \mathcal K=\overline{\{(x,d,\piid{t}(x)):x\in\cC,\ d\in\cD,\ t\in\bN\}} .
 \label{eq:nominalclosure}
\end{equation}
With the bounds below, $\mathcal K$ is compact.

\begin{assumption}[Linear independence of active rows]
\label{ass:licq}
At each $(x,d,z)\in\mathcal K$, the active rows of the oracle QP with reference $z$ are linearly independent, including active rows with zero multipliers.
\end{assumption}
For a time-invariant nominal controller this is pointwise LICQ on $\cC\times\cD$. Including the limiting reference commands gives a uniform regularity conclusion over time. At an optimizer with nonempty active set $\I$, the direction $-M_{\I}\tr(M_{\I}M_{\I}\tr)^{-1}\one_{|\I|}$ strictly decreases each active inequality. Taking a sufficiently short step gives a strictly feasible point because the remaining rows have slack. The empty active set already gives strict feasibility.

\begin{assumption}[Available model bounds]
\label{ass:bounds}
For all $x\in\cC$ and $t\in\bN$, known constants satisfy
\begin{align*}
 \norm{g(x)}&\leq\bar g, & \norm{Dg(x)}&\leq L_g,\\
 \norm{\piid{t}(x)}&\leq\bar\pi, & \norm{D\piid{t}(x)}&\leq L_\pi,\\
 \norm{\fd(x)}&\leq\bar f, & \norm{D\fd(x)}&\leq L_f,\\
 \norm{D F_{b,t}(x)}&\leq c, & F_{b,t}(x)&=\fd(x)+g(x)\piid{t}(x).
\end{align*}
\end{assumption}
Here $c$ is allowed to be at least one. If the filter always constrains every state direction, the relevant contraction test in Theorem~\ref{thm:certificate} uses the normal block alone. Define $\bar d=\max_{d\in\cD}\norm d$, $\bar H=\norm H$, $\bar E=\norm{E_u}$, and $\bar b=\big(\sum_i\norm{b_i}^2\big)^{1/2}$.

For a robust filter indexed by $j\in\{\mathrm{b},\mathrm{r}\}$, let $(x_t^j,u_t^j)$ and $(x_t^{\mathrm{o}},u_t^{\mathrm{o}})$ denote the respective robust and oracle closed-loop trajectories under the same disturbance sequence and initial state. The performance objective is to bound
\begin{equation}
 \Reg_j(T)=\sum_{t=0}^{T-1}
 \big[\ell_t(x_t^j,u_t^j)-\ell_t(x_t^{\mathrm{o}},u_t^{\mathrm{o}})\big].
 \label{eq:regret}
\end{equation}
The comparator is time dependent through $d_t$ and $\piid{t}$, so~\eqref{eq:regret} is a dynamic regret. Since both controllers apply safety filters to a given nominal command, this cost difference can have either sign.

\section{Robust Safe Control}
\label{sec:safe_control}

\subsection{Design with Baseline Barrier Functions}
\label{subsec:baseline}
At time $t$, let $u$ be a candidate virtual command. The prediction using the estimate and the actual successor are
\begin{align}
 \hat x_{t+1}(u)&=\fd(x_t)+g(x_t)u+P(x_t)\hat d_t,\\
 x_{t+1}(u)&=\hat x_{t+1}(u)+e_t .
 \label{eq:pred_actual}
\end{align}
The actual applied input is $u-g^\dagger(x_t)\hat d_t$. Equivalently, the actual successor is $\fd(x_t)+g(x_t)u+P(x_t)d_t+g(x_t)g^\dagger(x_t)e_t$. This form separates the unmatched disturbance from the error in compensating the matched component.

Let $B_i$ be Lipschitz continuous with constant $L_{B,i}$ on a set containing the predicted and actual successors. The baseline barrier condition is
\begin{equation}
 B_i(\hat x_{t+1}(u))\geq\alpha_i B_i(x_t)+L_{B,i} E_t,
 \quad i\in[N_B],
 \label{eq:baselinecondition}
\end{equation}
where $E_t$ is propagated using~\eqref{eq:estimation_error_bound}. This condition requires the estimate, its error bound, and the current state. For affine barriers, its left-hand side is affine in the command. Define
\begin{equation}
 r_t^{\mathrm b}=\begin{bmatrix}
     \col(L_{B,i} E_t)_{i\in[N_B]}\\ 0_{c_u}
 \end{bmatrix},\qquad
 \pi_{\mathrm b,t}(x)=\psi_t(x,\hat d_t,r_t^{\mathrm b}).
 \label{eq:baselineqp}
\end{equation}
The lower block of~\eqref{eq:data} enforces the physical input constraint after applying compensation.

\begin{proposition}[Baseline safety]
\label{prop:baseline}
Under Assumption~\ref{ass:variation}, suppose $E_0\geq\norm{e_0}$. If~\eqref{eq:baselinecondition} is enforced at every step, $B_i(x_{t+1})\geq\alpha_i B_i(x_t)$ for every $i$. In the affine case, the QP~\eqref{eq:baselineqp} is sufficient. With Assumption~\ref{ass:feasible} and $x_0\in\cC$, the resulting trajectory remains in $\cC$ and respects $\cU$.
\end{proposition}

Appendix~\ref{app:safety} proves the two safety results. The baseline margin is proportional to the uncertainty in the current estimate. Its limiting value under a constant increment bound is $L_{B,i}\bar\omega/(1-\decay)$. If a sharper certified sequence $\omega_t$ is available,~\eqref{eq:estimation_error_bound} reduces the margin as the past increments are discounted.

\subsection{Design with Robustified Barrier Functions}
\label{subsec:robustified}
To remove the need for $E_t$ in the online constraint, define the time-varying robustified barrier
\begin{equation}
 B_{\mathrm r,i,t}(x)=B_i(x)-\sigma_i\norm{e_t},\qquad
 \sigma_i=\frac{L_{B,i}}{\alpha_i-\decay},
 \label{eq:robustified_barrier}
\end{equation}
where the observer gain is chosen so that $\decay<\alpha_i$ for all $i$. We seek $B_{\mathrm r,i,t+1}(x_{t+1})\geq\alpha_i B_{\mathrm r,i,t}(x_t)$. The coefficient in~\eqref{eq:robustified_barrier} cancels the term in $\norm{e_t}$ when we bound this increment using~\eqref{eq:recursive_error}. The sufficient online condition is
\begin{equation}
 B_i(\hat x_{t+1}(u))\geq \alpha_i B_i(x_t)+\sigma_i\omega_t,
 \quad i\in[N_B].
 \label{eq:robustifiedcondition}
\end{equation}
For affine barriers, the corresponding QP is
\begin{equation}
 r_t^{\mathrm r}=\begin{bmatrix}\col(\sigma_i\omega_t)_{i\in[N_B]}\\0_{c_u}\end{bmatrix},
 \qquad \pi_{\mathrm r,t}(x)=\psi_t(x,\hat d_t,r_t^{\mathrm r}).
 \label{eq:robustifiedqp}
\end{equation}

\begin{proposition}[Safety with robustified barriers]
\label{prop:robustified}
Under Assumption~\ref{ass:variation}, let $\decay<\min_i\alpha_i$ and assume
\begin{equation}
 \norm{e_0}\leq\min_{i\in[N_B]}
 \frac{(\alpha_i-\decay)B_i(x_0)}{L_{B,i}}.
 \label{eq:initial_threshold}
\end{equation}
If~\eqref{eq:robustifiedcondition} is enforced at every step, then $B_{\mathrm r,i,t}(x_t)\geq0$ and $B_i(x_t)\geq0$ for all $i,t$. With affine barriers and Assumption~\ref{ass:feasible}, the filter~\eqref{eq:robustifiedqp} therefore keeps the state in $\cC$ and the applied input in $\cU$.
\end{proposition}

For an affine barrier, $B_i(x_0)/L_{B,i}$ is the signed distance to its defining hyperplane. Condition~\eqref{eq:initial_threshold} therefore relates the admissible initial error to the available safety slack and the rate gap $\alpha_i-\decay$. For an initial point on a barrier boundary it requires an exact initial estimate. It can be checked by substituting a known upper bound on $\norm{e_0}$, while the online QP requires only $\hat d_t$ and $\omega_t$. The robustified condition enforces decay of the augmented barrier and can permit the original barrier to decrease faster on a particular step. Its guarantee is forward invariance of the safe set from~\eqref{eq:initial_threshold}.

With $\omega_t=\bar\omega$, the online margin is $L_{B,i}\bar\omega/(\alpha_i-\decay)$. It is at least the limiting baseline margin, with equality when $\alpha_i=1$ or $\bar\omega=0$. If the baseline starts with $E_0>E_\infty$, the robustified margin can be smaller during the initial transient. In particular, with a constant disturbance and an exact observer update, its online margin is zero even during the transient. The two designs therefore trade the initial safety slack, the observer convergence rate, and the constraint tightening.

A squared penalty $B_i(x)-\sigma\norm{e_t}^2/2$ can also yield a sufficient online condition when $\decay^2<\alpha_i$. In this case a linear term in $\norm{e_t}$ remains in the barrier bound and is absorbed by completing the square, as detailed in Appendix~\ref{app:safety}. The norm penalty in~\eqref{eq:robustified_barrier} gives an exact cancellation in that bound and a margin proportional to $\omega_t$.

\section{Theoretical Analysis}
\label{sec:theoretical_analysis}

\subsection{Perturbation Analysis of Safety Filter}
\label{subsec:perturbation_analysis}
We compare the robust and oracle commands at the same state $x$ and time $t$. At this state the constraint matrix $M(x)$ and reference point $\piid{t}(x)$ are fixed. Only the right-hand side changes. This is the setting of polyhedral projection sensitivity~\cite{yen1995lipschitz,canovas2026lipschitz}. Define
\begin{equation}
 G(x)=\begin{bmatrix}HP(x)\\ E_u g^\dagger(x)\end{bmatrix},\qquad
 \bar G=\sqrt{\bar H^2+\bar E^2/\ug^2},
 \label{eq:Gperturbation}
\end{equation}
so that $\norm{G(x)}\leq\bar G$. For $j\in\{\mathrm b,\mathrm r\}$, the right-hand side perturbation is
\begin{equation}
 \delta\beta_t^j(x)=\beta(x,\hat d_t)-r_t^j-\beta(x,d_t)
                   =-G(x)e_t-r_t^j.
 \label{eq:rhsgap}
\end{equation}
The upper block includes the unmatched disturbance error. The lower block accounts for the change in the virtual command limits caused by compensation.

To characterize the sensitivity, consider the fixed-matrix projection map
\[
 \mathcal P_x(z,b)=\argmin_{M(x)u\leq b}\tfrac12\norm{u-z}^2
\]
at $z=\piid{t}(x)$ and $b=\beta(x,d_t)$. At this nominal point, LICQ implies strict feasibility. The formula of C\'anovas \textit{et al.}~\cite[eq.~(27)]{canovas2026lipschitz} gives the exact local modulus for joint perturbations of $(z,b)$ as
\begin{equation}
 \max_{\I\in\mathcal L}
 \norm{[\, I_m-M_{\I}^\dagger M_{\I}\quad M_{\I}^\dagger T_{\I}\,]}_{\mathrm{p}\to 2}.
 \label{eq:projection_modulus}
\end{equation}
The input norm for this formula is $\norm{(z,b)}_{\mathrm p}=\max\{\norm z,\norm{b}_\infty\}$, and the output norm is Euclidean. Here $T_{\I}b=b_{\I}$ and $\mathcal L$ consists of independent subsets $\I$ of the nominal active set whose row normals generate $z-\mathcal P_x(z,b)$ by a nonnegative linear combination. For the empty set this requires zero residual. Including all such subsets accounts for switches involving zero multipliers. In our comparison $z$ is fixed, so the relevant branch derivative is $M_{\I}^\dagger T_{\I}$. Its norm is $1/\sigmin(M_{\I})$ for nonempty $\I$ and zero for $\I=\varnothing$. For fixed $z$ and unrestricted Euclidean perturbations of the right-hand side, LICQ further gives the local modulus
\begin{equation}
 \operatorname{lip}_b\mathcal P_x(z,b)=
 \begin{cases}
  1/\sigmin(M_{\mathcal A_t(x,d_t)}(x)),& \mathcal A_t(x,d_t)\ne\varnothing,\\
  0,& \mathcal A_t(x,d_t)=\varnothing .
 \end{cases}
 \label{eq:rhs_modulus}
\end{equation}
Appendix~\ref{app:sensitivity} proves this expression, including weakly active rows. The particular perturbations in~\eqref{eq:rhsgap} may use only some right-hand side directions. To control a finite change, we need a bound that covers the entire perturbation, including changes of active rows away from the oracle solution.

\begin{proposition}[Sensitivity to the disturbance estimate and tightening]
\label{prop:sensitivity}
Under Assumptions~\ref{ass:domain}, \ref{ass:feasible}, \ref{ass:licq}, and \ref{ass:bounds}, there is a finite $\Lsf$ uniform over time and applicable along each robust trajectory such that
\begin{equation}
 \norm{\pi_{j,t}(x)-\pi_{\mathrm{o},t}(x,d_t)}
 \leq\Lsf\big(\bar G\norm{e_t}+\norm{r_t^j}\big).
 \label{eq:sensitivity}
\end{equation}
The same constant applies to any feasible comparison at $x\in\cC$ of the form~\eqref{eq:rhsgap}. If a uniform lower singular-value bound $\kappa>0$ is available for all nonempty independent KKT bases along the right-hand side segments joining the two problems, one can take $\Lsf=1/\kappa$.
\end{proposition}

Appendix~\ref{app:sensitivity} establishes the finite-perturbation bound. It uses a uniform neighborhood of the oracle data and compactness of the physical input set to cover larger changes. The segment estimate offers a direct numerical choice when the relevant row subsets have a certified singular-value bound. For the baseline, $\norm{r_t^{\mathrm b}}=\bar b E_t$ and $\bar G\leq\bar b+\bar E/\ug$. Hence~\eqref{eq:sensitivity} implies
\begin{equation}
 \norm{\pi_{\mathrm b,t}(x)-\pi_{\mathrm{o},t}(x,d_t)}
 \leq \Lsf(2\bar b+\bar E/\ug)E_t .
 \label{eq:sensitivity_baseline}
\end{equation}
For the robustified barrier, the tightening has norm $\big(\sum_i\sigma_i^2\big)^{1/2}\omega_t$. This term is separate from the contribution of the current estimation error, a distinction used in the regret analysis.

\subsection{Contraction Analysis of Safe Control Policies}
\label{subsec:contraction_analysis}
Contraction of the oracle closed loop gives an incremental estimate for states driven by the same disturbance sequence. With a bound on the effect of input perturbations, this yields the state deviation estimate used in Section~\ref{subsec:regret_analysis}, in the spirit of exponential incremental input-to-state stability~\cite{bayer2013discrete,liu2025regret}. We first derive the state Jacobian on each active set and then use it to obtain the uniform tests stated in Section~\ref{subsec:settings}. Throughout this subsection and the corresponding appendices, fix $t$ and write $\pi=\piid{t}$, $F_b=F_{b,t}$, $F=F_t$, and $\pi_{\mathrm o}=\pi_{\mathrm{o},t}$.

\paragraph{Active sets and derivatives}
Write $\I$ for an independent set of rows of $M(x)$. The KKT conditions for the projection are
\begin{equation}
 \begin{aligned}
  u-\pi+M\tr\lambda&=0,& Mu&\leq\beta,\\
  \lambda_i(M_i u-\beta_i)&=0,&\lambda&\geq0.
 \end{aligned}
 \label{eq:kkt}
\end{equation}
For a nonempty $\I$, the candidate branch and multiplier are
\begin{align}
 S_{\I}&=M_{\I}M_{\I}\tr, &
 L_{\I}&=M_{\I}\tr S_{\I}^{-1},\\
 \lambda_{\I}&=S_{\I}^{-1}(M_{\I}\pi-\beta_{\I}), &
 u_{\I}&=\pi-M_{\I}\tr\lambda_{\I},\\
 F_{\I}&=\fd+gu_{\I}+Pd .
 \label{eq:explicit}
\end{align}
For fixed $d$, the set on which $\I$ is exactly the set of active constraints is
\begin{equation}
 \cC_{\I}(d)=\left\{x\in\cC:
 \begin{array}{l}
 \rank M_{\I}(x)=|\I|,\quad \lambda_{\I}\geq0,\\
 M_{\I}u_{\I}=\beta_{\I},\quad
 M_{\I^c}u_{\I}<\beta_{\I^c}
 \end{array}\right\}.
 \label{eq:cell}
\end{equation}
On the empty-set branch $u_\varnothing=\pi$, $F_\varnothing=F_b+Pd$, and the cell is given by $M\pi<\beta$. An active row remains in the cell definition when its multiplier is zero. These cells are disjoint and cover $\cC$. They may be disconnected, and their closures may intersect.

\begin{proposition}[Branch regularity and state Jacobian]
\label{prop:branches}
Under Assumptions~\ref{ass:domain}, \ref{ass:feasible}, \ref{ass:licq}, and \ref{ass:bounds}, the oracle optimizer and state map are continuous and locally Lipschitz in $(x,d)$ and locally select from finitely many $C^1$ maps~\eqref{eq:explicit}. The applied oracle input is uniformly Lipschitz in $x$ on $\cC$, with some $L_{\mathrm o}<\infty$ independent of $t$ and $d$. For $V_{\I}^u=I_m-L_{\I}M_{\I}$ and any direction $v$,
\begin{align}
 Du_{\I}[v]
 &= V_{\I}^u\big(D\pi[v]-(DM_{\I}[v])\tr\lambda_{\I}\big)
 \notag\\
 &\quad+L_{\I}\big(D\beta_{\I}[v]-(DM_{\I}[v])u_{\I}\big),
 \label{eq:Du}\\
 DF_{\I}[v]
 &=D\fd[v]+Dg[v]u_{\I}+gDu_{\I}[v]+DP[v]d,\label{eq:DF}\\
 DM[v]&=\begin{bmatrix}-HDg[v]\\0\end{bmatrix},\notag\\
 D\beta[v]&=
 \begin{bmatrix}
 H(D\fd[v]+DP[v]d)-A_\alpha Hv\\
 E_u Dg^\dagger[v]d
 \end{bmatrix}.
 \label{eq:Ddata}
\end{align}
If $\norm{DF_{\I}(x,d)}\leq q$ on the closure of each cell, uniformly in $t,d$, then $\Lip_x(F_t(\cdot,d);\cC)\leq q$.
\end{proposition}

The proof in Appendix~\ref{app:branches} gives the pseudoinverse and projector derivatives and establishes the global Lipschitz estimate across switches, including switches on cell boundaries. The lower block of $D\beta[v]$ accounts for the motion of the physical input limits in the virtual input coordinates.

\paragraph{Geometry of the active equalities}
Split the active set into barrier indices $\B\subseteq[N_B]$ and physical input indices $\J\subseteq[c_u]$ with $r=|\B|$ and $\ell=|\J|$. Write $H_{\B}$ and $E_{\J}$ for the corresponding rows of $H$ and $E_u$. The orthogonal projector onto $\ker E_{\J}$ is $V_{\J}=I_m-E_{\J}^\dagger E_{\J}$, with $V_\varnothing=I_m$. Let $\Phi_{\J}$ be the state-update map after projecting the nominal physical input onto the equalities $E_{\J}w=\one_\ell$, and $J_{\J}=D_x\Phi_{\J}$. The remaining correction enforces the active barrier equalities. For $r>0$ it uses
\begin{equation}
 \begin{aligned}
 W_{\J}&=gV_{\J}g\tr,&
 K_{\I}&=W_{\J}H_{\B}\tr(H_{\B}W_{\J}H_{\B}\tr)^{-1},\\
 A_{\B}&=\diag(\alpha_i)_{i\in\B},&
 t_{\B}&=A_{\B}H_{\B}x+(I_r-A_{\B})s_{\B}.
 \end{aligned}
 \label{eq:geodef_main}
\end{equation}
The inverse exists under LICQ. The exact branch map is $F_{\I}=\Phi_{\J}+K_{\I}(t_{\B}-H_{\B}\Phi_{\J})$. In orthonormal coordinates for the tangent and normal spaces of the active barriers, its Jacobian has the form
\begin{equation}
 \begin{gathered}
 Q_{\B}J_{\I}Q_{\B}\tr
   =\begin{bmatrix}\mathcal A_{\I}&\mathcal B_{\I}\\0&D_{\B}\end{bmatrix},\\
 D_{\B}=L_{\B}^{-1}A_{\B}L_{\B},\quad
 L_{\B}=(H_{\B}H_{\B}\tr)^{1/2}.
 \end{gathered}
 \label{eq:tri_main}
\end{equation}
Here $J_{\I}=D_xF_{\I}$ and $Q_{\B}$ is constructed in Appendix~\ref{app:geometry}. The normal block follows from $H_{\B}F_{\I}=t_{\B}$. The tangent block depends on the physical input equalities and on the derivative of $K_{\I}$ when the actuation varies with the state. With common retention, $D_{\B}=\alpha I_r$. With unequal retention, its induced norm can exceed the largest $\alpha_i$ because the barrier normals need not be orthogonal. The induced norm of the full Jacobian governs discrete-time Euclidean contraction.

\paragraph{Constants for the uniform tests}
For each fixed matrix $E_u$ and $H$, enumerate its nonempty, independent row subsets of size at most $m$. Let $e_*$ and $h_*$ be the smallest singular values over these finite families for $E_u$ and $H$, respectively. If a family is empty, the associated active class is absent and its constants can be omitted. For an active barrier block,
\begin{equation}
 d_* = \max_{\substack{\varnothing\ne\B,\ |\B|\leq m\\
                       \rank H_{\B}=|\B|}}
 \norm{(H_{\B}H_{\B}\tr)^{-1/2}A_{\B}(H_{\B}H_{\B}\tr)^{1/2}} .
 \label{eq:dglobal}
\end{equation}
The simpler bound $d_*\leq(\bar H/h_*)\alpha_{\max}$ is also valid, where $\alpha_{\max}=\max_i\alpha_i$. With common retention $d_*=\alpha$. All of these extrema may use a smaller family if an activity exclusion is verified from the QP data.

Let $L_\dagger$ and $L_P$ bound the derivatives of $g^\dagger$ and $P$, respectively. From Appendix~\ref{app:branches}, valid choices are
\begin{equation}
 (L_\dagger,L_P)=
 \begin{cases}
 (L_g/\ug^2,0), & m=n,\\
 (2L_g/\ug^2,2L_g/\ug), & m<n .
 \end{cases}
 \label{eq:LdagLP}
\end{equation}
If the column space of $g$ is constant, use $L_P=0$. Define
\begin{align}
 \Delta_*&=\bar\pi+\bar d/\ug+\sqrt m/e_* ,\label{eq:Delta}\\
 C_U^{(1)}&=c+L_g\Delta_*+\bar gL_\pi+(L_P+\bar gL_\dagger)\bar d,\notag\\
 C_U^{(2)}&=L_f+L_g\Delta_*+\bar gL_\pi+\bar gL_\dagger\bar d,\notag\\
 C_U&=\min\{C_U^{(1)},C_U^{(2)}\},\qquad C_0=c+L_P\bar d,\notag\\
 C_*&=\max\{C_0,C_U\}.
 \label{eq:Cglobal}
\end{align}
The bounds $C_0$ and $C_U$ apply to $J_{\J}$ for $\J=\varnothing$ and $\J\ne\varnothing$, respectively. If physical limits are never active, one may use $C_*=C_0$ and $\Delta_*=\bar\pi+\bar d/\ug$. A tighter bound on $J_{\J}$ obtained from the known model can also be used. If $L_f$ is unavailable, $L_f=c+L_g\bar\pi+\bar gL_\pi$ is valid by differentiating $\fd=F_b-g\pi$. Write $R_x=\max_{x\in\cC}\norm x$ and $\alpha_{\min}=\min_i\alpha_i$. The residual in the barrier correction has the bound
\begin{multline}
 R_*=\bar H\big(\alpha_{\max}R_x+\bar f
                  +\bar g\Delta_*+\bar d\big)\\
                    +(1-\alpha_{\min})\norm s .
 \label{eq:Rglobal}
\end{multline}

The projection in~\eqref{eq:geodef_main} is governed by the actuation directions left by the active physical limits. To bound them, choose a reference $x_{\rm ref}\in\cC$ and set
\begin{equation}
 G_0=g(x_{\rm ref}),\quad
 R_0=\max_{x\in\cC}\norm{x-x_{\rm ref}},\quad
 \delta_g=L_gR_0 .
 \label{eq:gvariation}
\end{equation}
A sharper certified bound on $\norm{g(x)-G_0}$ can replace $\delta_g$. Construct a finite family $\mathfrak F$ of pairs $(\B,\J)$ that contains all active patterns. One can include every pair with $|\B|+|\J|\leq m$ and with $H_{\B}$ and $E_{\J}$ of full row rank when nonempty. An identically dependent stacked row set is excluded by LICQ. Other exclusions require verifying that the pattern cannot occur. Rank deficiency only at the reference state does not justify exclusion.

When partial barrier activity is possible, use the following singular-value calculation:
\begin{align}
 \eta_0&=\min_{\substack{(\B,\J)\in\mathfrak F\\0<|\B|<n}}
                   \sigmin(H_{\B}G_0V_{\J}),\\
 \eta&=\eta_0-\bar H\delta_g>0 .
 \label{eq:eta}
\end{align}
A positive $\eta$ bounds the minimum singular value of $H_{\B}g(x)V_{\J}$ on every retained partial pattern. In the special case $g(x)=a(x)G_0$ with $a(x)\geq a_{\min}>0$, use $\eta=a_{\min}\eta_0$. Each corresponding $K_{\I}$ is then constant. Define
\begin{equation}
 \begin{gathered}
 p_* = \max\{1,\bar g\bar H/\eta\},\quad
 \chi_* = \sqrt{p_*^2-1},\\
 \eps_* = 2p_*\bar g L_g\bar H R_* /\eta^2 .
 \end{gathered}
 \label{eq:globalgeometry}
\end{equation}
Here $p_*$ bounds the norm of the projection onto the free state directions, $\chi_*$ bounds the coupling into those directions, and $\eps_*$ bounds the term induced by the variation of $K_{\I}$. For constant $K_{\I}$, take $\eps_*=0$. A nonpositive $\eta$ leaves the reference test inconclusive. This test uses fixed row-subset calculations and known bounds on $g$, with no need to locate the active-set cells. The bound on the directional geometry is essential when an actuator limit and a barrier can become active simultaneously, as shown in Appendix~\ref{app:certificate}. Table~\ref{tab:certificate_data} summarizes the data required by the test.

\begin{table}[t]
\caption{Known data for the uniform contraction test.}
\label{tab:certificate_data}
\centering
\small
\setlength{\tabcolsep}{3pt}
\renewcommand{\arraystretch}{1.16}
\begin{tabular}{@{}p{0.23\columnwidth}p{0.73\columnwidth}@{}}
\hline
Quantity & Data and calculation \\
\hline
$e_*,h_*$ & Singular values of independent row subsets of $E_u$ and $H$. \\
$d_*$ & Matrix norms in~\eqref{eq:dglobal} using $H$ and $\alpha_i$. \\
$C_*$ & Model bounds, $\bar d$, and $e_*$ in~\eqref{eq:Cglobal}. \\
$R_*$ & Model bounds and safe-set data in~\eqref{eq:Rglobal}. \\
$\eta$ & Reference actuation, the constraint rows, and $\delta_g$ in~\eqref{eq:eta}. \\
$p_*,\chi_*,\eps_*$ & Geometry margin and model bounds in~\eqref{eq:globalgeometry}. \\
\hline
\end{tabular}
\end{table}

For the square-actuation corollaries, $L_W$ bounds the derivative of $W=gg\tr$, and $2\bar g L_g$ is one such bound. For branches without active physical input limits define
\begin{align}
 R_b&=\bar H(\alpha R_x+\bar f+\bar g\bar\pi)+(1-\alpha)\norm s ,\notag\\
 \chi_b&=\frac{\bar w-\underline w}{2\sqrt{\underline w\bar w}},
 \quad p_b=\frac{\bar w+\underline w}{2\sqrt{\underline w\bar w}},
 \quad \eps_b=\frac{p_bL_W\bar H R_b}{\underline w h_*^2}.
 \label{eq:conditionedconstants}
\end{align}
Set $q_U=0$ if physical limits are never active and $q_U=C_U$ if input-active branches have no active barriers. If mixed activity is possible, set
\begin{equation}
 q_U=\sqrt{(p_*C_U+\chi_*\alpha+\eps_*)^2+\alpha^2}.
 \label{eq:qU}
\end{equation}
These cases refer to activity possible over the oracle domain. The geometry test for~\eqref{eq:qU} only needs the mixed patterns with $\J\ne\varnothing$ and $0<r<n$. If the $\alpha_i$ differ, the square-actuation bounds use $d_*$ in the normal-block terms and $\alpha_{\max}$ and $1-\alpha_{\min}$ in $R_b$. Under any of the resulting contraction tests, two oracle trajectories driven by the same disturbance sequence have state difference at most $q^t\norm{x_0-y_0}$. A fixed nominal controller and a constant disturbance also give a unique fixed point in $\cC$ by the contraction mapping theorem.

\subsection{Regret Analysis of Robust Safe Control}
\label{subsec:regret_analysis}
Suppose the oracle closed loop is uniformly contractive with factor $q<1$, for example by Theorem~\ref{thm:certificate}. Following the trajectory comparison in~\cite{liu2025regret}, we treat the difference between the robust and oracle filters at the same state as an input perturbation of the oracle dynamics. This separates the accumulated effect on the state from the immediate difference in the applied command.

Write $w_{j,t}(x)=\pi_{j,t}(x)-g^\dagger(x)\hat d_t$ and $F_{j,t}(x)=\fd(x)+g(x)w_{j,t}(x)+d_t$, holding the estimate and margin of the robust trajectory fixed in these expressions. Subtracting the oracle dynamics gives
\begin{align}
 F_{j,t}(x)-F_t(x,d_t)
 &=g(x)\big(\pi_{j,t}(x)-\pi_{\mathrm{o},t}(x,d_t)\big)\notag\\
 &\quad+g(x)g^\dagger(x)e_t .
 \label{eq:dynamics_gap}
\end{align}
The matched projector in the second term has norm at most one. In the input difference the corresponding term is $g^\dagger(x)e_t$. Define
\begin{align}
 a_u&=\Lsf\bar G+\ug^{-1}, & b_u&=\Lsf,\\
 a_x&=\bar g\Lsf\bar G+1, & b_x&=\bar g\Lsf,
 \label{eq:perturbation_constants}
\end{align}
and let $\eta_{u,t}=a_u\norm{e_t}+b_u\norm{r_t^j}$ and $\eta_{x,t}=a_x\norm{e_t}+b_x\norm{r_t^j}$. These bound the input and state-update differences at $x=x_t^j$. The oracle applied input also has a uniform Lipschitz bound $L_{\mathrm o}$ in the state, by Proposition~\ref{prop:branches}. Equation~\eqref{eq:oracle_input_constant} gives a numerical choice when the oracle active rows have certified conditioning. Write
\begin{align}
 L_c&=L_{\ell,x}+L_{\ell,u}L_{\mathrm o},\\
 K_e(q)&=L_{\ell,u}a_u+L_c a_x/(1-q),\\
 K_r(q)&=L_{\ell,u}b_u+L_c b_x/(1-q).
 \label{eq:regret_constants}
\end{align}

\begin{theorem}[Finite-time dynamic regret]
\label{thm:regret}
Under Assumptions~\ref{ass:cost_lipschitz}, \ref{ass:domain}, \ref{ass:variation}, \ref{ass:feasible}, \ref{ass:licq}, and \ref{ass:bounds}, suppose the oracle is uniformly contractive with factor $q<1$. For the baseline and robustified filters, initialized as in Propositions~\ref{prop:baseline} and~\ref{prop:robustified}, respectively, and compared from the same state, the absolute cost gap satisfies
\begin{align}
 |\Reg_j(T)|
 &\leq L_{\ell,u}\sum_{t=0}^{T-1}\eta_{u,t}
   +L_c\sum_{t=0}^{T-2}\Gamma_{T-1-t}(q)\eta_{x,t}
 \label{eq:regret_finite}\\
 &\leq K_e(q)\sum_{t=0}^{T-1}\norm{e_t}
       +K_r(q)\sum_{t=0}^{T-1}\norm{r_t^j}.
 \label{eq:regret_general}
\end{align}
\end{theorem}

The first bound accounts for the remaining horizon at each step. In particular, the last input perturbation contributes to the last stage cost directly, and its effect on $x_T$ falls outside the cost in~\eqref{eq:regret}. The proof in Appendix~\ref{app:regret} uses contraction only for the oracle and sums the resulting geometric propagation of state errors.

For $\Omega_T=\sum_{t=0}^{T-1}\omega_t$, summing the observer envelope gives
\begin{align}
 \sum_{t=0}^{T-1}E_t
 &=\Gamma_T(\decay) E_0
   +\sum_{t=0}^{T-2}\Gamma_{T-1-t}(\decay)\omega_t \notag\\
 &\leq \frac{E_0+\Omega_T}{1-\decay}.
 \label{eq:sum_envelope}
\end{align}
The same estimate holds for the sum of $\norm{e_t}$ with the actual initial error in place of $E_0$. For constant $\omega_t=\bar\omega$, the first line equals $(E_0-E_\infty)\Gamma_T(\decay)+T E_\infty$. For the robustified filter let $S_\sigma=\big(\sum_i\sigma_i^2\big)^{1/2}\leq\bar b/(\alpha_{\min}-\decay)$. Inserting the two margins into~\eqref{eq:regret_general} yields
\begin{align}
 |\Reg_{\mathrm b}(T)|&\leq
 (K_e(q)+\bar b K_r(q))\,\frac{E_0+\Omega_T}{1-\decay},\\
 |\Reg_{\mathrm r}(T)|&\leq
 K_e(q)\,\frac{\norm{e_0}+\Omega_T}{1-\decay}
 +K_r(q) S_\sigma\Omega_T .
 \label{eq:design_regret}
\end{align}
The actual initial error appears in the analysis of the second filter and is bounded by~\eqref{eq:initial_threshold}. One can sharpen the observer contribution after the disturbance and innovation sequences have been realized. Define $V_T=\sum_{t=0}^{T-1}\norm{\Delta_{d,t}}$ and $N_T=\sum_{t=0}^{T-1}\norm{\Lambda\sn{t}}$. Then~\eqref{eq:observer_exact} gives
\begin{equation}
 \sum_{t=0}^{T-1}\norm{e_t}
 \leq\frac{\norm{e_0}+V_T+N_T}{1-\decay}.
 \label{eq:realized_variation}
\end{equation}
This quantity can replace the factor multiplying $K_e(q)$ in each design bound, while the term multiplying $K_r(q)$ is determined by the tightening actually applied. This distinction matters when the realized variation is small but the filters use a conservative online envelope.

With constant envelopes, $\Omega_T=T(\bar\Delta_d+\lambda_{\max}(\Lambda)\sno)$. Holding the other model and sensitivity constants fixed, the baseline bound scales as
\begin{equation}
 O\!\left(\frac{E_0+T(\bar\Delta_d+\lambda_{\max}(\Lambda)\sno)}
                  {(1-q)(1-\decay)}\right).
 \label{eq:scaling_base}
\end{equation}
For the robustified design the observer term has the same dependence with $\norm{e_0}$ in place of $E_0$, and the margin adds
\begin{equation}
 O\!\left(\frac{T(\bar\Delta_d+\lambda_{\max}(\Lambda)\sno)}
                  {(1-q)(\alpha_{\min}-\decay)}\right).
 \label{eq:scaling_robustified}
\end{equation}
The denominators distinguish propagation by the plant from error decay in the observer and the rate gap needed by the robustified barrier. A faster observer can reduce the accumulation due to disturbance increments, while the innovation term also depends on the largest observer gain.

For fixed $q$ and $\Lambda$, both designs have $o(T)$ regret if the envelopes used in the filters satisfy $\Omega_T=o(T)$. Summable envelopes yield bounded regret. In particular, a constant disturbance with exact innovation and a zero increment bound gives a bounded initialization cost. If the filters use a fixed positive worst-case bound, the corresponding tightening term in the certificate grows linearly with $T$, even when the realized disturbances vary slowly. These conclusions concern the effect of disturbance information and robustification for the same nominal controller.

\section{Conclusions and future work}
\label{sec:conclusion}
We developed two observer-based safety filters for discrete-time nonlinear systems with additive disturbances. The baseline barrier condition propagates a bound on the disturbance estimation error. The robustified barrier condition trades that online bound for a requirement on the observer convergence rate and a safety-dependent initialization condition. With affine barriers, both designs use QPs that enforce the physical input limits after compensation.

We bounded the effect of estimation and tightening on the QP solution and established sufficient conditions for uniform contraction of the oracle closed loop. The contraction test accounts for the actuation directions available after imposing active physical limits. These results give dynamic regret bounds in terms of the initial error, the cumulative disturbance variation and observer innovation error, and the contraction factor. They also show when the margins used by the filters permit sublinear regret. The fixed-state sensitivity result also applies to continuous-time CBF filters whose uncertainty changes the right-hand side of a QP with a fixed constraint matrix. Future work will examine less conservative contraction tests for systems with widely varying actuation geometry and extend the performance analysis to disturbances that depend on the state or input.

\section*{Acknowledgment}
This work is part of a project funded by the European Research Council (ERC) under the European Union's Horizon 2020 research and innovation programme (Grant Agreement No. 101018826, CLariNet).

\bibliographystyle{IEEEtran}
\bibliography{4-references/references}
\appendices
\section{Observer and Safety Proofs}
\label{app:observer}

\paragraph{Observer error}
Subtracting two consecutive estimation errors gives
\begin{align*}
 e_{t+1}-e_t
 &= (d_{t+1}-\hat d_{t+1})-(d_t-\hat d_t)\notag\\
 &=\Delta_{d,t}-\Lambda(x_{t+1}-x_t)+(\xi_{t+1}-\xi_t)\notag\\
 &=\Delta_{d,t}-\Lambda(f(x_t,u_t)+d_t-x_t)\notag\\
 &\quad+\Lambda(f(x_t,u_t)-x_t+\hat d_t+\sn{t})\notag\\
 &=\Delta_{d,t}-\Lambda e_t+\Lambda\sn{t}.
\end{align*}
Adding $e_t$ gives~\eqref{eq:recursive_error}. Since $\Lambda$ is symmetric, write $\Lambda=Q\Sigma Q\tr$ with $Q$ orthogonal. Then $A_\Lambda=Q(I_n-\Sigma)Q\tr$ and $A_\Lambda^j=Q(I_n-\Sigma)^j Q\tr$ for each $j\in\bN$. Successive substitution gives
\begin{equation}
 e_t=A_\Lambda^t e_0+
 \sum_{k=0}^{t-1}A_\Lambda^{t-1-k}(\Delta_{d,k}+\Lambda\sn{k}).
 \label{eq:observer_exact}
\end{equation}
As $\norm{Q}=\norm{Q\tr}=1$ and $\norm{I_n-\Sigma}=\decay$, the triangle inequality gives
\begin{align*}
 \norm{e_t}
 &\leq\decay^t\norm{e_0}+\sum_{k=0}^{t-1}
   \decay^{t-1-k}(\norm{\Delta_{d,k}}+\norm{\Lambda\sn{k}})\\
 &\leq\decay^t E_0+\sum_{k=0}^{t-1}\decay^{t-1-k}\omega_k .
\end{align*}
With the constant bound $\bar\omega$ the sum is $\bar\omega(1-\decay^t)/(1-\decay)$, giving the expression below~\eqref{eq:estimation_error_bound}. For $\sn{t}=0$ the same bound has $E_\infty=\bar\Delta_d/(1-\decay)$ when the constant increment bound is used.

\paragraph{Proof of Proposition~\ref{prop:baseline}}
\label{app:safety}
By~\eqref{eq:pred_actual}, the Lipschitz condition yields
\begin{align*}
 B_i(x_{t+1})
 &= B_i(\hat x_{t+1})+
       \big(B_i(x_{t+1})-B_i(\hat x_{t+1})\big)\notag\\
 &\geq B_i(\hat x_{t+1})-L_{B,i}\norm{e_t}\notag\\
 &\geq\alpha_i B_i(x_t)+L_{B,i}(E_t-\norm{e_t})\notag\\
 &\geq\alpha_i B_i(x_t).
\end{align*}
Induction gives $B_i(x_t)\geq\alpha_i^tB_i(x_0)\geq0$. In the affine case, expanding~\eqref{eq:baselinecondition} gives
\begin{multline*}
 b_i\tr(\fd(x_t)+g(x_t)u+P(x_t)\hat d_t)-s_i\\
 \geq\alpha_i(b_i\tr x_t-s_i)+L_{B,i}E_t .
\end{multline*}
Rearranging gives
\begin{multline*}
 -b_i\tr g(x_t)u\leq b_i\tr(\fd(x_t)+P(x_t)\hat d_t)\\
                       -\tau_i(x_t)-L_{B,i}E_t .
\end{multline*}
This is the $i$th upper row of~\eqref{eq:baselineqp}. The lower block is equivalent to $E_u(u-g^\dagger(x_t)\hat d_t)\leq\one_{c_u}$. The safe set includes all required state bounds, so induction and feasibility give the result. \hfill\qedsymbol

\paragraph{Proof of Proposition~\ref{prop:robustified}}
The required barrier increment is nonnegative if
\begin{align*}
 B_i(x_{t+1}) &\geq
 \alpha_i B_i(x_t)+\sigma_i(\norm{e_{t+1}}-\alpha_i\norm{e_t}).
\end{align*}
Using the prediction instead gives the sufficient condition
\begin{multline*}
 B_i(\hat x_{t+1})\geq \alpha_i B_i(x_t)+L_{B,i}\norm{e_t}\\
                    +\sigma_i(\norm{e_{t+1}}-\alpha_i\norm{e_t}).
\end{multline*}
The error recursion implies
\begin{align*}
 \norm{e_{t+1}}-\alpha_i\norm{e_t}
 &=\norm{A_\Lambda e_t+\Delta_{d,t}+\Lambda\sn{t}}
        -\alpha_i\norm{e_t}\notag\\
 &\leq(\decay-\alpha_i)\norm{e_t}+\omega_t .
\end{align*}
The total error term in the sufficient condition is at most
\[
 \big[L_{B,i}+\sigma_i(\decay-\alpha_i)\big]\norm{e_t}
                 +\sigma_i\omega_t .
\]
The choice of $\sigma_i$ in~\eqref{eq:robustified_barrier} makes the coefficient zero. This proves sufficiency of~\eqref{eq:robustifiedcondition}. Rearranging~\eqref{eq:initial_threshold} gives $B_{\mathrm r,i,0}(x_0)\geq0$. Hence
\[
 B_i(x_t)\geq B_{\mathrm r,i,t}(x_t)
            \geq\alpha_i^t B_{\mathrm r,i,0}(x_0)\geq0 .
\]
Expanding the affine condition as in the baseline proof replaces $L_{B,i}E_t$ in the upper row by $\sigma_i\omega_t$. The physical constraints are again equivalent to the lower block of~\eqref{eq:data}. \hfill\qedsymbol

For the squared penalty discussed in Section~\ref{subsec:robustified}, write $v=\norm{e_t}$ and $a=\alpha_i-\decay^2>0$. Then $\norm{e_{t+1}}^2\leq(\decay v+\omega_t)^2$, so the error term in the sufficient condition satisfies
\begin{align*}
 &L_{B,i}v+\tfrac\sigma2(\norm{e_{t+1}}^2-\alpha_i v^2)\\
 &\quad\leq -\tfrac{\sigma a}{2}v^2
       +(L_{B,i}+\sigma\decay\omega_t)v+\tfrac\sigma2\omega_t^2\\
 &\quad=-\tfrac{\sigma a}{2}
 \left(v-\frac{L_{B,i}+\sigma\decay\omega_t}{\sigma a}\right)^2\\
 &\qquad\quad+\tfrac\sigma2\omega_t^2+
   \frac{(L_{B,i}+\sigma\decay\omega_t)^2}{2\sigma a}.
\end{align*}
The last two terms give a margin independent of $v$. Initialization requires $B_i(x_0)\geq\sigma\norm{e_0}^2/2$. For zero $\omega_t$ and fixed $\sigma>0$, this bound still uses a positive margin when $L_{B,i}>0$.

\section{Regularity and Perturbation of the Safety Filter}
\label{app:branches}

\paragraph{Solving and differentiating the KKT equations}
A feasible convex QP with affine inequalities satisfies the KKT equations because the normal cone to its feasible polyhedron is the cone generated by the active row normals. Multipliers on inactive rows vanish. A minimal conical representation of the normal yields an independent set supporting stationarity, called a KKT basis. Under Assumption~\ref{ass:licq}, the entire active set is independent and supports stationarity with nonnegative multipliers, including zeros if needed. For this set $\I$, the active equations are
\[
 u_{\I}-\pi+M_{\I}\tr\lambda_{\I}=0,\qquad M_{\I}u_{\I}=\beta_{\I}.
\]
Substitution gives $M_{\I}\pi-S_{\I}\lambda_{\I}=\beta_{\I}$ and then~\eqref{eq:explicit}. Because $S_{\I}$ is invertible, the formulas extend to a $C^1$ map on a neighborhood where the same rank holds. Conversely, on~\eqref{eq:cell} the candidate satisfies the KKT conditions with exactly those active rows and therefore is the unique optimizer.

Suppress $\I$ in the derivatives of the active equations. Let $r=M\pi-\beta$ and $S=MM\tr$. Then
\begin{align*}
 \lambda&=S^{-1}r,\\
 DS[v]&=(DM[v])M\tr+M(DM[v])\tr,\\
 Dr[v]&=(DM[v])\pi+M D\pi[v]-D\beta[v].
\end{align*}
Differentiating $S\lambda=r$ gives
\begin{align}
 D\lambda[v]=S^{-1}\big(& (DM[v])\pi+M D\pi[v]-D\beta[v]\notag\\
 &-\big((DM[v])M\tr+M(DM[v])\tr\big)\lambda\big).
 \label{eq:Dlambda}
\end{align}
Since $u=\pi-M\tr\lambda$,
\[
 Du[v]=D\pi[v]-(DM[v])\tr\lambda-M\tr D\lambda[v].
\]
Using $\pi-M\tr\lambda=u$ in~\eqref{eq:Dlambda} yields
\begin{multline*}
 D\lambda[v]=S^{-1}\big(M(D\pi[v]-(DM[v])\tr\lambda)\\
                           -D\beta[v]+(DM[v])u\big).
\end{multline*}
Substituting gives~\eqref{eq:Du}. The product rule gives~\eqref{eq:DF}, and differentiating the two blocks of~\eqref{eq:data} gives~\eqref{eq:Ddata}.

For the pseudoinverse derivative, put $N=g\tr g$. Then $DN[v]=(Dg[v])\tr g+g\tr Dg[v]$, and
\begin{align}
 Dg^\dagger[v]
 &=-N^{-1}(DN[v])N^{-1}g\tr+N^{-1}(Dg[v])\tr\notag\\
 &=-g^\dagger Dg[v]g^\dagger+N^{-1}(Dg[v])\tr P .
 \label{eq:Ddag}
\end{align}
Differentiating $P=I_n-gg^\dagger$ and using $gN^{-1}=(g^\dagger)\tr$ gives
\begin{align}
 DP[v]&=-Dg[v]g^\dagger-gDg^\dagger[v]\notag\\
      &=-P Dg[v]g^\dagger-(g^\dagger)\tr(Dg[v])\tr P .
 \label{eq:DP}
\end{align}
For unit $v$, the first term of~\eqref{eq:Ddag} is bounded by $L_g/\ug^2$ and the second by $L_g/\ug^2$ because $\norm{N^{-1}}\leq1/\ug^2$ and $\norm P\leq1$. Each term in~\eqref{eq:DP} is bounded by $L_g/\ug$. For square actuation $P=0$ and $DP=0$, so only the first term in~\eqref{eq:Ddag} remains. This proves~\eqref{eq:LdagLP}.

\paragraph{Continuity under changes of the QP data}
Consider the map $U(x,z,b)=\mathcal P_x(z,b)$ introduced in Section~\ref{subsec:perturbation_analysis} and fix a point $(x_0,z_0,b_0)$ with oracle data. Here $x_0,z_0,b_0$ are generic parameters in this proof. Let $u_0=U(x_0,z_0,b_0)$. The direction in Section~\ref{sec:problem_formulation} gives a strictly feasible $\tilde u_0$, which remains feasible for all data sufficiently close to $(x_0,z_0,b_0)$ by continuity. For a convergent sequence $(x_k,z_k,b_k)$ of such data, the associated minimizers $u_k$ are uniformly bounded. Indeed, if the physical block of $b_k$ is bounded above by $a\one_{c_u}$ with $a>0$, then $E_u u_k\leq a\one_{c_u}$ implies $u_k\in a\cU$.

The limit $\bar u$ of any convergent subsequence of $u_k$ is feasible for the limit data. For any $v$ feasible at the limit, $v_\eps=(1-\eps)v+\eps\tilde u_0$ with $0<\eps\leq1$ is strictly feasible there and feasible for all sufficiently large $k$. Optimality gives
\[
 \tfrac12\norm{u_k-z_k}^2\leq\tfrac12\norm{v_\eps-z_k}^2 .
\]
Taking the subsequence limit and then $\eps\downarrow0$ shows that $\bar u$ minimizes the limit QP. Uniqueness implies $\bar u=u_0$ and hence continuity. On oracle data,~\eqref{eq:F} gives continuity of the oracle state map and applied input.

\paragraph{Local regularity and the segment argument}
Rows that are slack at $(x_0,z_0,b_0)$ remain slack for all sufficiently close data. The active set at any such point is therefore a subset of the nominal active set. Every subset has full row rank in a small neighborhood because the nominal rows are independent. For each subset,~\eqref{eq:explicit} with $z,b$ in place of $\pi,\beta$ defines a $C^1$ map on that neighborhood. The actual optimizer is a continuous selection of these finitely many maps. Such selections are locally Lipschitz~\cite[Ch.~4]{scholtes2012}. The same conclusion also follows from strong regularity of the KKT system~\cite{robinson1980strongly}: LICQ holds at the nominal data and the objective Hessian is $I_m$, so the strong second-order sufficient condition holds.

To pass from branch derivatives to a global bound for $F$, fix $t$ and $d$ and parametrize the segment $\gamma(a)=x+a(y-x)$ for $0\leq a\leq1$. Convexity keeps the segment in $\cC$. Local Lipschitz continuity and compactness of the segment make $a\mapsto F(\gamma(a),d)$ absolutely continuous. At almost every $a$ the derivative exists. Take a sequence of nearby points on the segment converging to $\gamma(a)$. By local finiteness a subsequence uses a single branch $\I$. Continuity gives equality of the branch and the oracle map at $\gamma(a)$. Difference quotients along the subsequence give
\[
 \frac{d}{da}F(\gamma(a),d)=DF_{\I}(\gamma(a),d)(y-x).
\]
The limiting point is in the closure of that branch cell. If the derivative bound in Proposition~\ref{prop:branches} holds, integration gives
\[
 \norm{F(y,d)-F(x,d)}
 \leq\int_0^1 q\norm{y-x}\,da=q\norm{y-x}.
\]
Absolute continuity permits infinitely many switches along the segment.

To bound the oracle input uniformly, the image of $\mathcal K$ in the full QP data $(x,z,b)$, with $b=\beta(x,d)$, is compact. The local Lipschitz neighborhoods for $U$ on this set have a finite subcover, with a common upper bound on their Lipschitz constants. Along each oracle segment the QP data $x$, $z=\piid{t}(x)$ and $b=\beta(x,d)$ are uniformly Lipschitz with respect to $x$ by Assumption~\ref{ass:bounds} and~\eqref{eq:Ddata}--\eqref{eq:LdagLP}. The applied input also subtracts $g^\dagger(x)d$, with derivative bounded by $L_\dagger\bar d$. Subdividing each segment into pieces in the local neighborhoods and adding the bounds gives a uniform $L_{\mathrm o}$ for the applied input. For example, if $L_U$ is the common local Lipschitz bound on $U$ in the Euclidean product norm, one valid choice is
\[
 L_{\mathrm o}= L_U\sqrt{1+L_\pi^2+L_\beta^2}+L_\dagger\bar d ,
\]
where $L_\beta$ bounds $\norm{D_x\beta}$ uniformly. One can obtain $L_\beta$ by bounding the two blocks of~\eqref{eq:Ddata} using Assumption~\ref{ass:bounds} and~\eqref{eq:LdagLP}. This completes the proof of Proposition~\ref{prop:branches}. \hfill\qedsymbol

\paragraph{Proof of Proposition~\ref{prop:sensitivity}}
\label{app:sensitivity}
First consider~\eqref{eq:rhs_modulus}, with $x$ and $z$ fixed. Write $\mathcal A=\mathcal A_t(x,d_t)$. If $\mathcal A$ is empty, $z$ stays strictly feasible for all sufficiently small changes in $b$, so the local modulus is zero. Otherwise, only subsets of $\mathcal A$ can become active locally. The Gram matrix for any nonempty subset $\I\subseteq\mathcal A$ is a principal submatrix of $M_{\mathcal A}M_{\mathcal A}\tr$. Its least eigenvalue is at least that of the full Gram matrix, by evaluating the quadratic form on vectors supported on $\I$. Hence
\[
 \norm{M_{\I}^\dagger T_{\I}}
 \leq 1/\sigmin(M_{\mathcal A}).
\]
Integrating the corresponding branch slopes along local right-hand side segments gives the upper bound in~\eqref{eq:rhs_modulus}.

For the reverse bound, let $u_0=\mathcal P_x(z,b)$ and let $\lambda_{\mathcal A}$ be its multiplier on $\mathcal A$. For $\eps>0$ define
\begin{align*}
 \lambda^\eps_{\mathcal A}&=\lambda_{\mathcal A}+\eps\one_{|\mathcal A|},\\
 u^\eps&=z-M_{\mathcal A}\tr\lambda^\eps_{\mathcal A},\\
 b^\eps_{\mathcal A}&=M_{\mathcal A}u^\eps,\\
 b^\eps_{\mathcal A^c}&=b_{\mathcal A^c}.
\end{align*}
As $\eps\downarrow0$, these variables converge to their nominal values. For small $\eps$, the inactive rows stay slack and the active multipliers are strictly positive. The KKT conditions then give the same active set throughout an open neighborhood of $b^\eps$. The derivative on this neighborhood is $M_{\mathcal A}^\dagger T_{\mathcal A}$, of norm $1/\sigmin(M_{\mathcal A})$. Since these neighborhoods occur arbitrarily close to $b$, the local modulus has at least this value.

For the finite perturbation in~\eqref{eq:rhsgap}, use the compact set of oracle data from the preceding argument. By shrinking each local neighborhood before taking a finite subcover, there are a common radius $\delta_0>0$ and constant $L_U$ such that, whenever $\norm{\delta\beta_t^j(x)}\leq\delta_0$, the two QP data points lie in a common neighborhood with that Lipschitz bound. Hence
\[
 \norm{\pi_{j,t}(x)-\pi_{\mathrm{o},t}(x,d_t)}
 \leq L_U\norm{\delta\beta_t^j(x)} .
\]
For a larger perturbation, the two physical inputs belong to $\cU$, so the virtual-command difference is at most
\[
 \diam(\cU)+\norm{g^\dagger(x)e_t}.
\]
Since $\cU$ is bounded, $E_u$ has full column rank. Otherwise a nonzero vector in $\ker E_u$ would generate an unbounded line of feasible inputs. The lower block of~\eqref{eq:rhsgap} is $-E_u g^\dagger(x)e_t$. Consequently,
\[
 \norm{g^\dagger(x)e_t}
 \leq\norm{E_u^\dagger}\norm{E_u g^\dagger(x)e_t}
 \leq\norm{E_u^\dagger}\norm{\delta\beta_t^j(x)}.
\]
Dividing the diameter term by $\delta_0$ yields the uniform constant
\[
 \Lsf=\max\{L_U,\diam(\cU)/\delta_0+\norm{E_u^\dagger}\}.
\]
The triangle inequality in~\eqref{eq:rhsgap} yields~\eqref{eq:sensitivity}.

For the segment estimate, fix $M(x)$ and $z$ and let the right-hand side interpolate affinely between the two values. Feasibility along the segment follows by interpolating feasible endpoint commands. For any independent KKT basis, the candidate optimizer is affine in the interpolation parameter by~\eqref{eq:explicit}, with derivative $M_{\I}^\dagger T_{\I}\delta\beta_t^j(x)$. The portion of the segment where a given basis satisfies the KKT conditions is an interval, since feasibility and multiplier nonnegativity are affine inequalities in that parameter. The finite family of intervals covers the segment because an independent basis exists at every optimizer, and the branches agree on overlaps by uniqueness. Adding the estimates on these intervals gives the bound with $\Lsf=1/\kappa$. \hfill\qedsymbol

The reference-matrix calculation of Section~\ref{subsec:contraction_analysis} can also bound the row conditioning in this last argument if its candidate family covers all such KKT bases. In a mixed pattern write $B=H_{\B}gV_{\J}$ and suppose $\sigmin(B)\geq\eta$. To solve $M_{\I}v=\col(b_B,b_U)$, take $v_0=E_{\J}^\dagger b_U$ and $v=v_0+B^\dagger(-b_B-H_{\B}gv_0)$. The second term lies in $\ker E_{\J}$. Thus
\[
 \norm v\leq\eta^{-1}\norm{b_B}
      +e_*^{-1}(1+\bar H\bar g/\eta)\norm{b_U}.
\]
This bounds the norm of the minimum-norm right inverse since $M_{\I}^\dagger$ produces the minimum-norm solution. For pure actuator patterns the bound is $e_*^{-1}$. For pure barrier patterns use the bound for $H_{\B}g$ directly. When $r=n$ and $m=n$, $H_{\B}$ and $g$ are invertible and $\sigmin(H_{\B}g)\geq h_*\ug$. More generally, with no active physical limit, the square-actuation condition $gg\tr\succeq\underline w I_n$ gives $\sigmin(H_{\B}g)\geq h_*\sqrt{\underline w}$ for every nonempty $\B$.

\paragraph{An explicit oracle input Lipschitz bound}
Suppose $\nu$ is a uniform upper bound on $\norm{M_{\I}^\dagger}$ for the nonempty oracle active sets. The row-conditioning estimates above give one way of computing $\nu$ from known data. Write $U_0=\max_{w\in\cU}\norm w+\bar d/\ug$ and $L_M=\bar H L_g$. Then $\norm{u_{\I}}\leq U_0$ on the oracle cells and $\norm{DM_{\I}}\leq L_M$. The stationarity equation gives
\[
 \lambda_{\I}=(M_{\I}^\dagger)\tr(\pi-u_{\I}),
 \qquad \norm{\lambda_{\I}}\leq\nu(\bar\pi+U_0).
\]
Equation~\eqref{eq:Ddata} also gives the derivative bound
\begin{multline*}
 L_\beta=\big[\bar H^2(L_f+L_P\bar d+\alpha_{\max})^2
                       +\bar E^2L_\dagger^2\bar d^2\big]^{1/2}.
\end{multline*}
Using $\norm{V_{\I}^u}\leq1$ in~\eqref{eq:Du}, and adding the derivative of the compensation term, yields
\begin{multline}
 L_{\mathrm o}=L_\pi+L_M\nu(\bar\pi+U_0)
        +\nu(L_\beta+L_M U_0)\\
        +L_\dagger\bar d .
 \label{eq:oracle_input_constant}
\end{multline}
This value also bounds the empty-set branch. The same segment argument as for $F$ establishes the bound for the applied oracle input across switches.

\section{Active-Set Geometry and Triangularization}
\label{app:geometry}

Split a selected active set into barrier rows \(\B\) and actuator
rows \(\J\), with \(r=|\B|\), \(\ell=|\J|\). Write
\[
 H_{\B}=\col(b_i\tr)_{i\in\B},\qquad E_{\J}=(E_u)_{\J},
 \qquad M_{\I}=\begin{bmatrix}-H_{\B}g\\E_{\J}\end{bmatrix}.
\]
For nonempty \(\J\), define the following constant matrices and
state-dependent maps:
\begin{equation}
\begin{split}
 E_{\J}^\dagger&=E_{\J}\tr(E_{\J}E_{\J}\tr)^{-1},\qquad
 V_{\J}=I_m-E_{\J}^\dagger E_{\J},\\
 w_{\J}^0&=V_{\J}(\pi-g^\dagger d)+E_{\J}^\dagger\one_\ell,\\
 u_{\J}^0&=w_{\J}^0+g^\dagger d
       =V_{\J}\pi+E_{\J}^\dagger\one_\ell
                         +(I_m-V_{\J})g^\dagger d .
\end{split}
\label{eq:inputprojection}
\end{equation}
For empty \(\J\), set \(V_{\J}=I_m\),
\(w_{\J}^0=\pi-g^\dagger d\), and \(u_{\J}^0=\pi\).
The map \(w_{\J}^0\) is the projection of \(\rho=\pi-g^\dagger d\) onto the selected
\emph{physical} equalities \(E_{\J}w=\one_\ell\).
Set
\begin{equation}
 \begin{gathered}
 \Phi_{\J}=\fd+g w_{\J}^0+d=\fd+g u_{\J}^0+Pd,\\
 W_{\J}=gV_{\J}g\tr,\qquad J_{\J}=D_x\Phi_{\J}.
 \end{gathered}
 \label{eq:Phi}
\end{equation}
For \(r>0\), also set
\begin{equation}
\begin{split}
 A_{\B}&=\diag(\alpha_i)_{i\in\B},\quad
 t_{\B}=A_{\B}H_{\B}x+(I_r-A_{\B})s_{\B},\\
 G_{\I}&=H_{\B}W_{\J}H_{\B}\tr,\quad
 K_{\I}=W_{\J}H_{\B}\tr G_{\I}^{-1},\\
 e_{\I}&=t_{\B}-H_{\B}\Phi_{\J}.
\end{split}
\label{eq:geodef}
\end{equation}
Full row rank of \(M_{\I}\) implies \(r+\ell\leq m\),
full row rank of \(H_{\B},E_{\J}\), and \(G_{\I}\succ0\).

\begin{lemma}[Explicit dynamics and the projection derivative]
\label{lem:geometry}
For \(r>0\), the physical input and state-update maps are
\begin{align}
 w_{\I}&=w_{\J}^0+V_{\J}g\tr H_{\B}\tr G_{\I}^{-1}e_{\I},
 \label{eq:wbranch}\\
 u_{\I}&=w_{\I}+g^\dagger d,\qquad
 F_{\I}=\Phi_{\J}+K_{\I}e_{\I}.       \label{eq:Fgeo}
\end{align}
Their state Jacobian is
\begin{equation}
 \begin{gathered}
 J_{\I}=(I_n-K_{\I}H_{\B})J_{\J}+K_{\I}A_{\B}H_{\B}+R_{\I},\\
 R_{\I}v=DK_{\I}[v]e_{\I},
 \end{gathered}                              \label{eq:Jgeo}
\end{equation}
where
\begin{align}
 DK_{\I}[v]
 &=(I_n-K_{\I}H_{\B})DW_{\J}[v]H_{\B}\tr G_{\I}^{-1},
 \label{eq:DK}\\
 DW_{\J}[v]&=Dg[v]V_{\J}g\tr+gV_{\J}(Dg[v])\tr,\qquad
 H_{\B}R_{\I}=0.                                     \label{eq:DW}
\end{align}
For \(r=0\), simply \(F_{\I}=\Phi_{\J}\) and \(J_{\I}=J_{\J}\).
\end{lemma}

\begin{proof}
For a chosen set of active rows, the optimizer also minimizes
the objective subject to its active equalities alone: the KKT
stationarity equation proves this, since its Hessian is \(I_m\).
Let \(\rho=\pi-g^\dagger d\). Use the physical-input representation \eqref{eq:physicalqp}.
The projection onto \(E_{\J}w=\one_\ell\) is \(w_{\J}^0\).
Any other input satisfying these equalities is \(w=w_{\J}^0+z\),
where \(E_{\J}z=0\). Since
\(w_{\J}^0-\rho\in\range E_{\J}\tr\), orthogonality gives
\[
 \norm{w-\rho}^2
    =\norm{w_{\J}^0-\rho}^2+\norm z^2 .
\]
It remains to find the minimum-norm \(z\in\ker E_{\J}\)
satisfying the active barrier equalities:
\[
 H_{\B}gz=t_{\B}-H_{\B}(\fd+gw_{\J}^0+d)=e_{\I}.
\]
Let \(B=H_{\B}gV_{\J}\). The minimum-norm solution of \(Bz=e_{\I}\)
is \(B\tr(BB\tr)^{-1}e_{\I}\).
It lies in \(\range V_{\J}=\ker E_{\J}\), so it also solves
the constrained minimum-norm problem above.
Since \(BB\tr=G_{\I}\), this proves \eqref{eq:wbranch}.
Multiplying the correction by \(g\) proves \eqref{eq:Fgeo}.

For completeness, the Gram matrix of the full active rows is
\[
 M_{\I}M_{\I}\tr=
 \begin{bmatrix}
 H_{\B}gg\tr H_{\B}\tr&-H_{\B}gE_{\J}\tr\\
 -E_{\J}g\tr H_{\B}\tr&E_{\J}E_{\J}\tr
 \end{bmatrix}.
\]
Its Schur complement after eliminating \(E_{\J}E_{\J}\tr\) is
\(H_{\B}gV_{\J}g\tr H_{\B}\tr=G_{\I}\), which is positive
definite by LICQ. This also verifies the required rank of \(B\).

Differentiate \(F_{\I}=\Phi_{\J}+K_{\I}(t_{\B}-H_{\B}\Phi_{\J})\):
\[
 DF_{\I}[v]=(I-K_{\I}H_{\B})J_{\J}v
             +K_{\I}A_{\B}H_{\B}v+DK_{\I}[v]e_{\I}.
\]
Next \(DG_{\I}[v]=H_{\B}DW_{\J}[v]H_{\B}\tr\), so
\[
 \begin{split}
 DK_{\I}[v]
 &=DW_{\J}[v]H_{\B}\tr G_{\I}^{-1}\\
 &\quad -W_{\J}H_{\B}\tr G_{\I}^{-1}
      (H_{\B}DW_{\J}[v]H_{\B}\tr)G_{\I}^{-1}\\
 &=(I-K_{\I}H_{\B})DW_{\J}[v]H_{\B}\tr G_{\I}^{-1}.
 \end{split}
\]
Finally \(H_{\B}K_{\I}=I_r\). Differentiating this identity
gives \(H_{\B}DK_{\I}[v]=0\), and hence \(H_{\B}R_{\I}=0\).
\end{proof}

The input-equality Jacobian can be written
in either of two equivalent ways:
\begin{align}
 J_{\J}v
 &=D\fd[v]+Dg[v]w_{\J}^0\notag\\
 &\quad+gV_{\J}\bigl(D\pi[v]-Dg^\dagger[v]d\bigr),       \label{eq:Jphysical}\\
 &=DF_b[v]+DP[v]d+Dg[v](u_{\J}^0-\pi)\notag\\
 &\quad-g(I_m-V_{\J})D\pi[v]
       +g(I_m-V_{\J})Dg^\dagger[v]d.                    \label{eq:Jnominal}
\end{align}
For \(\J=\varnothing\), the latter reduces to
\(J_{\J}v=DF_b[v]+DP[v]d\).
Both expressions include the state-dependent shift associated with compensation.

\paragraph{Orthogonal triangular form}
For \(r>0\), let \(L_{\B}=(H_{\B}H_{\B}\tr)^{1/2}\),
\(Y_{\B}=H_{\B}\tr L_{\B}^{-1}\).
Choose \(Z_{\B}\) with orthonormal columns spanning \(\ker H_{\B}\).
Then
\begin{equation}
 Q_{\B}=\begin{bmatrix}Z_{\B}\tr\\Y_{\B}\tr\end{bmatrix}
 \quad\text{is orthogonal}.
\end{equation}
It is constant on each branch because it depends only on the fixed barrier normals.
Define
\begin{equation}
 C_{\I}=Z_{\B}\tr K_{\I}L_{\B},\qquad
 D_{\B}=L_{\B}^{-1}A_{\B}L_{\B}.
 \label{eq:CD}
\end{equation}

\begin{theorem}[Exact triangularization]
\label{thm:triangular}
For every branch with \(r>0\),
\begin{equation}
 Q_{\B}J_{\I}Q_{\B}\tr
  =\begin{bmatrix}\mathcal A_{\I}&\mathcal B_{\I}\\0&D_{\B}\end{bmatrix}.
 \label{eq:tri}
\end{equation}
If \(Q_{\B}J_{\J}Q_{\B}\tr=[J_{ab}]_{a,b=1}^2\) and
\(Q_{\B}R_{\I}Q_{\B}\tr=
\left[\begin{smallmatrix}R_{11}&R_{12}\\0&0\end{smallmatrix}\right]\),
then
\begin{equation}
 \begin{aligned}
 \mathcal A_{\I}&=J_{11}-C_{\I}J_{21}+R_{11},\\
 \mathcal B_{\I}&=J_{12}-C_{\I}J_{22}+C_{\I}D_{\B}+R_{12}.
 \end{aligned}
 \label{eq:AB}
\end{equation}
For common retention \(\alpha_i=\alpha\), \(D_{\B}=\alpha I_r\).
For \(r=n\), \(Q_{\B}J_{\I}Q_{\B}\tr=D_{\B}\) exactly.
\end{theorem}
\begin{proof}
The active barrier equalities imply
\(H_{\B}F_{\I}=t_{\B}\), hence \(H_{\B}J_{\I}=A_{\B}H_{\B}\).
Since \(H_{\B}Z_{\B}=0\), \(H_{\B}Y_{\B}=L_{\B}\), the lower-left
block is zero, and the lower-right block equals
\(L_{\B}^{-1}A_{\B}L_{\B}\).
To calculate the remaining blocks, put \(\Pi_{\I}=K_{\I}H_{\B}\).
One has
\begin{align*}
 Q_{\B}\Pi_{\I}Q_{\B}\tr&=
 \begin{bmatrix}0&C_{\I}\\0&I_r\end{bmatrix},\\
 Q_{\B}(I_n-\Pi_{\I})Q_{\B}\tr&=
 \begin{bmatrix}I_{n-r}&-C_{\I}\\0&0\end{bmatrix}.
\end{align*}
Furthermore the right column of
\(Q_{\B}K_{\I}A_{\B}H_{\B}Q_{\B}\tr\) is
\(\col(C_{\I}D_{\B},D_{\B})\) and its left column is zero.
Insert these identities into \eqref{eq:Jgeo} to obtain
\eqref{eq:AB}. For \(r=n\) the upper blocks have zero dimension.
\end{proof}

\section{Derivation of the Contraction Certificates}
\label{app:certificate}

\paragraph{Fixed matrices and the input-equality bounds}
Each nonempty active set has at most $m$ rows. For any independent subsets of $E_u$ and $H$ that can occur,
\begin{equation}
 \begin{aligned}
  \norm{E_{\J}}&\leq\norm{E_u},&\sigmin(E_{\J})&\geq e_*,\\
  \norm{H_{\B}}&\leq\bar H,&\sigmin(H_{\B})&\geq h_*.
 \end{aligned}
 \label{eq:subbounds}
\end{equation}
The full matrices may have very different conditioning from their row subsets. For instance, two nearly parallel rows can give a small singular value in an otherwise well-conditioned matrix. The extrema in~\eqref{eq:subbounds} use each independent subset. The factors in the normal block satisfy
\[
 \norm{L_{\B}}=\norm{H_{\B}}\leq\bar H,\qquad
 \norm{L_{\B}^{-1}}=1/\sigmin(H_{\B})\leq1/h_* .
\]
Thus $\norm{D_{\B}}\leq\bar H\alpha_{\max}/h_*$. For common retention, $D_{\B}=\alpha I_r$, which gives the exact value $\alpha$ for every nonempty normal block.

The function-value bounds in Assumption~\ref{ass:bounds} can be obtained from reference values and $R_0$ in~\eqref{eq:gvariation}. Integrating each derivative along the segment from $x_{\rm ref}$ to $x$ gives the valid choices
\begin{align*}
 \bar f&=\norm{\fd(x_{\rm ref})}+L_fR_0,\\
 \bar\pi&=\sup_t\norm{\piid{t}(x_{\rm ref})}+L_\pi R_0,\\
 \bar g&=\norm{g(x_{\rm ref})}+L_gR_0 .
\end{align*}
For a polytope, radii defined by the maximum of a norm are obtained by maximizing the same norm over its vertices.

For nonempty $\J$, the projection formulas in~\eqref{eq:inputprojection} give
\begin{align*}
 \norm{w_{\J}^0}
 &\leq \norm{V_{\J}}\big(\norm\pi+\norm{g^\dagger d}\big)
       +\norm{E_{\J}^\dagger}\norm{\one_\ell}\notag\\
 &\leq \bar\pi+\bar d/\ug+\sqrt m/e_*=\Delta_*,\\
 u_{\J}^0-\pi&=-(I_m-V_{\J})\pi+E_{\J}^\dagger\one_\ell
                          +(I_m-V_{\J})g^\dagger d .
\end{align*}
Because $\norm{I_m-V_{\J}}=1$ for nonempty $\J$, this also gives $\norm{u_{\J}^0-\pi}\leq\Delta_*$. For the empty set, $\norm{w_\varnothing^0}\leq\bar\pi+\bar d/\ug\leq\Delta_*$. In~\eqref{eq:Jnominal}, the five terms for a unit direction are bounded, respectively, by $c$, $L_P\bar d$, $L_g\Delta_*$, $\bar g L_\pi$, and $\bar g L_\dagger\bar d$. Their sum gives $C_U^{(1)}$. In~\eqref{eq:Jphysical}, the bounds on the drift term, the derivative of $g$ times $w_{\J}^0$, and the two projected terms give $L_f$, $L_g\Delta_*$, $\bar gL_\pi$, and $\bar gL_\dagger\bar d$. Their sum gives $C_U^{(2)}$. With $\J=\varnothing$, $J_{\J}=DF_b+DP[\cdot]d$, giving $C_0$. This proves~\eqref{eq:Cglobal}.

The residual $e_{\I}=t_{\B}-H_{\B}\Phi_{\J}$ is bounded using
\begin{align*}
 \norm{\Phi_{\J}}&\leq\bar f+\bar g\Delta_*+\bar d,\\
 \norm{t_{\B}}&\leq\alpha_{\max}\bar H R_x+
                          (1-\alpha_{\min})\norm s .
\end{align*}
Since $\norm{H_{\B}}\leq\bar H$, this yields $\norm{e_{\I}}\leq R_*$ in~\eqref{eq:Rglobal}.

\paragraph{Directionality and the reference geometry bound}
Consider fixed rows $H_{\B}=[1\ 0]$, $E_{\J}=[1\ 0]$, and $V_{\J}=\diag(0,1)$. For a rotation angle $\vartheta\in(0,\pi/2)$ let
\[
 g_\vartheta=\begin{bmatrix}
 \cos\vartheta&-\sin\vartheta\\
 \sin\vartheta&\cos\vartheta
 \end{bmatrix}.
\]
The stacked rows are independent, $g_\vartheta g_\vartheta\tr=I_2$, $\norm{g_\vartheta}=\sigmin(g_\vartheta)=1$, and the gain is constant in the state. Yet $H_{\B}g_\vartheta V_{\J}=[0\ -\sin\vartheta]$ has smallest singular value $\sin\vartheta$. Here $K_{\I}=[1\ -\cot\vartheta]\tr$, so the tangent-normal coupling grows unboundedly as $\vartheta\downarrow0$. Thus the full gain's singular values alone cannot bound the geometry after imposing active physical limits.

Convexity of $\cC$ and the derivative bound on $g$ yield $\norm{g(x)-G_0}\leq L_g R_0=\delta_g$. For each retained partially barrier-active pattern, the singular-value perturbation inequality gives
\begin{align*}
 \sigmin(H_{\B}g(x)V_{\J})
 &\geq\sigmin(H_{\B}G_0V_{\J})\\
 &\quad-\norm{H_{\B}(g(x)-G_0)V_{\J}}\\
 &\geq\eta_0-\bar H\delta_g=\eta.
\end{align*}
A positive lower bound therefore follows from finitely many reference SVDs. If $g(x)=a(x)G_0$ with $a(x)\geq a_{\min}>0$, the left-hand side equals $a(x)\sigmin(H_{\B}G_0V_{\J})$. In addition, $W_{\J}$ and $G_{\I}$ both scale by $a(x)^2$, so the scaling cancels in $K_{\I}$. This proves the structural alternative in Section~\ref{subsec:contraction_analysis}.

Write $B=H_{\B}gV_{\J}$ for the remaining barrier actuation. Then $G_{\I}=BB\tr$ and $K_{\I}=gV_{\J}B^\dagger$, giving
\[
 \norm{K_{\I}H_{\B}}\leq\bar g\bar H/\eta .
\]
The orthogonal coordinates in Appendix~\ref{app:geometry} give, for $0<r<n$,
\begin{equation}
 \norm{K_{\I}H_{\B}}^2
 =\norm{I_n-K_{\I}H_{\B}}^2=1+\norm{C_{\I}}^2.
 \label{eq:projection_norm}
\end{equation}
Indeed, the nonzero column block of $Q_{\B}K_{\I}H_{\B}Q_{\B}\tr$ is $\col(C_{\I},I_r)$, whose squared norm is $1+\norm{C_{\I}}^2$. The nonzero row block of the complementary projection is $[I_{n-r}\ -C_{\I}]$ with the same squared norm. This gives the first two bounds in~\eqref{eq:globalgeometry}. From~\eqref{eq:DW}, $\norm{DW_{\J}[v]}\leq2\bar gL_g\norm v$. Also $\norm{G_{\I}^{-1}}\leq1/\eta^2$. Equation~\eqref{eq:DK} therefore gives
\[
 \norm{DK_{\I}[v]e_{\I}}
 \leq p_*\,(2\bar gL_g\norm v)\,\bar H\,\eta^{-2}R_*
 =\eps_*\norm v .
\]

\paragraph{Proof of Theorem~\ref{thm:certificate}}
If $r=0$, then $J_{\I}=J_{\J}$ and $\norm{J_{\I}}\leq C_*$ by~\eqref{eq:Cglobal}. If $r=n$, the only block of the triangularized Jacobian is $D_{\B}$, so $\norm{J_{\I}}\leq d_*$. For $0<r<n$, using~\eqref{eq:AB}, the upper block row satisfies
\begin{align*}
 \norm{[\,\mathcal A_{\I}\ \mathcal B_{\I}\,]}
 &\leq \norm{[\,I_{n-r}\ -C_{\I}\,]}\norm{J_{\J}}
             +\norm{C_{\I}D_{\B}}+\norm{R_{\I}}\notag\\
 &\leq p_*C_*+\chi_*d_*+\eps_*=:t_* .
\end{align*}
For $z=\col(z_1,z_2)$ the exact triangularization then gives
\begin{align*}
 \norm{Q_{\B}J_{\I}Q_{\B}\tr z}^2
 &\leq t_*^2\norm z^2+d_*^2\norm{z_2}^2\notag\\
 &\leq(t_*^2+d_*^2)\norm z^2 .
\end{align*}
Thus $\norm{J_{\I}}\leq q_*$. As $p_*\geq1$, this bound is also at least $C_*$ and $d_*$ and covers the other two cases. All the bounds apply on the closures of the relevant cells. Proposition~\ref{prop:branches} then gives the uniform contraction claim. Taking only the bounds for classes that can occur proves the reductions. \hfill\qedsymbol

If a barrier with $\alpha_i=1$ is active on a full-dimensional region, the normal block has eigenvalue one and precludes strict contraction there. If the barrier is never active, its row may be excluded from the normal-block calculation. In contrast, when the active barriers span the state on every branch, the test depends only on the normal block and can succeed even with an expansive nominal state map.

\paragraph{Proof of Corollary~\ref{cor:conditioned}}
On a branch with $\J=\varnothing$, use the orthogonal coordinates $[Z_{\B}\ Y_{\B}]$ from Appendix~\ref{app:geometry}. With $B=Y_{\B}\tr WY_{\B}$ and $H_{\B}=L_{\B}Y_{\B}\tr$, direct substitution gives
\[
 C_{\I}=Z_{\B}\tr WY_{\B}B^{-1}.
\]
The spectrum of $W$ lies in $[\underline w,\bar w]$, so
\[
 W^2\preceq(\underline w+\bar w)W-\underline w\bar w I_n .
\]
The lower-right block in the orthogonal coordinates yields
\begin{multline*}
 (Z_{\B}\tr WY_{\B})\tr(Z_{\B}\tr WY_{\B})+B^2\\
 \preceq(\underline w+\bar w)B-\underline w\bar w I_r .
\end{multline*}
Multiply on both sides by $B^{-1}$ to obtain
\[
 C_{\I}\tr C_{\I}\preceq
 (\underline w+\bar w)B^{-1}-\underline w\bar w B^{-2}-I_r .
\]
For each positive $z$ the scalar polynomial satisfies
\[
 (\underline w+\bar w)z-\underline w\bar w z^2-1
 \leq\frac{(\bar w-\underline w)^2}{4\underline w\bar w}.
\]
Applying this inequality to the eigenvalues of $B^{-1}$ proves $\norm{C_{\I}}\leq\chi_b$. Equation~\eqref{eq:projection_norm} then gives $\norm{I_n-K_{\I}H_{\B}}\leq p_b$, using $1+\chi_b^2=p_b^2$.

Here $\Phi_\varnothing=F_b$ because $P=0$ and compensation is exact in the state map. Consequently $\norm{e_{\I}}\leq R_b$. Also $G_{\I}=H_{\B}WH_{\B}\tr\succeq\underline w h_*^2 I_r$. Substituting these bounds and $\norm{DW[v]}\leq L_W\norm v$ into~\eqref{eq:DK} yields $\norm{R_{\I}}\leq\eps_b$. The upper-row argument for Theorem~\ref{thm:certificate} with $J_\varnothing=DF_b$ gives the displayed $q_b$. When $\J\ne\varnothing$, an input-only branch is bounded by $C_U$. A mixed branch has $r<n$ because $r+\ell\leq n$, so the partial-activity argument using $C_U$ in place of $C_*$ gives the third case in~\eqref{eq:qU}. That case also bounds input-only branches because $p_*\geq1$. Taking the maximum over the two physical input classes proves~\eqref{eq:corfull}. \hfill\qedsymbol

\paragraph{Proof of Corollary~\ref{cor:isotropic}}
For $W=a(x)I_n$ and $\J=\varnothing$, the scalar $a(x)$ cancels and $K_{\I}=H_{\B}\tr(H_{\B}H_{\B}\tr)^{-1}$. The projection is orthogonal, so $C_{\I}=0$ and the term $DK_{\I}[v]e_{\I}$ vanishes. Thus $p_b=1$, $\chi_b=0$, and $\eps_b=0$ can be used on these branches. Substitution gives $q_b=\sqrt{c^2+\alpha^2}$. The other class still uses $q_U$, since active physical equalities replace $W$ by $W_{\J}=gV_{\J}g\tr$. The pure classes give $q_b=\max\{c,\alpha\}$ when partial barrier activity is excluded. \hfill\qedsymbol

\section{Proof of the Dynamic Regret Bounds}
\label{app:regret}
At a common state $x=x_t^j$, subtract the applied oracle input from the robust input:
\begin{align*}
 w_{j,t}(x)-w_{\mathrm{o},t}(x,d_t)
 &=\pi_{j,t}(x)-\pi_{\mathrm{o},t}(x,d_t)+g^\dagger(x)e_t .
\end{align*}
Proposition~\ref{prop:sensitivity} and $\norm{g^\dagger}\leq1/\ug$ give
\begin{align*}
 \norm{w_{j,t}(x)-w_{\mathrm{o},t}(x,d_t)}
 &\leq\Lsf(\bar G\norm{e_t}+\norm{r_t^j})+\ug^{-1}\norm{e_t}\notag\\
 &=\eta_{u,t}.
\end{align*}
Similarly, multiplying the virtual-command difference by $g(x)$ and using $\norm{gg^\dagger}\leq1$ in~\eqref{eq:dynamics_gap} gives
\begin{align*}
 \norm{F_{j,t}(x)-F_t(x,d_t)}
 &\leq\bar g\Lsf(\bar G\norm{e_t}+\norm{r_t^j})+\norm{e_t}\notag\\
 &=\eta_{x,t}.
\end{align*}
This uses the common exogenous disturbance $d_t$ in the two plant transitions.

Let $z_t=\norm{x_t^j-x_t^{\mathrm{o}}}$. Inserting the oracle map at the robust state and using the contraction factor $q$ gives
\begin{align*}
 z_{t+1}
 &\leq\norm{F_{j,t}(x_t^j)-F_t(x_t^j,d_t)}\notag\\
 &\quad+\norm{F_t(x_t^j,d_t)-F_t(x_t^{\mathrm{o}},d_t)}\notag\\
 &\leq\eta_{x,t}+qz_t .
\end{align*}
Repeated substitution yields
\[
 z_t\leq q^t z_0+\sum_{k=0}^{t-1}q^{t-1-k}\eta_{x,k}.
\]
Here $z_0=0$ by the common initialization. The applied oracle input is Lipschitz in state by Proposition~\ref{prop:branches}. Inserting its value at $x_t^j$ gives
\begin{align*}
 \norm{u_t^j-u_t^{\mathrm{o}}}
 &\leq\norm{w_{j,t}(x_t^j)-w_{\mathrm{o},t}(x_t^j,d_t)}\notag\\
 &\quad+\norm{w_{\mathrm{o},t}(x_t^j,d_t)-w_{\mathrm{o},t}(x_t^{\mathrm{o}},d_t)}\notag\\
 &\leq\eta_{u,t}+L_{\mathrm o}z_t .
\end{align*}
Hence Assumption~\ref{ass:cost_lipschitz} yields
\begin{align*}
 \big|\ell_t(x_t^j,u_t^j)-\ell_t(x_t^{\mathrm{o}},u_t^{\mathrm{o}})\big|
 &\leq L_{\ell,x}z_t+L_{\ell,u}(\eta_{u,t}+L_{\mathrm o}z_t)\notag\\
 &= L_c z_t+L_{\ell,u}\eta_{u,t}.
\end{align*}
Summing the state terms and reversing the order of summation gives
\begin{align*}
 \sum_{t=0}^{T-1} z_t
 &\leq\sum_{t=1}^{T-1}\sum_{k=0}^{t-1}q^{t-1-k}\eta_{x,k}\\
 &=\sum_{k=0}^{T-2}\eta_{x,k}\sum_{t=k+1}^{T-1}q^{t-1-k}\\
 &=\sum_{k=0}^{T-2}\Gamma_{T-1-k}(q)\eta_{x,k}.
\end{align*}
The triangle inequality for the sum of stage-cost differences proves~\eqref{eq:regret_finite}. Since $\Gamma_{T-1-k}(q)\leq(1-q)^{-1}$, substitution of $\eta_{u,t},\eta_{x,t}$ and collecting the coefficients gives~\eqref{eq:regret_general}. With different initial states the same derivation would add $L_c\Gamma_T(q)z_0$ to~\eqref{eq:regret_finite}. \hfill\qedsymbol

To derive~\eqref{eq:sum_envelope}, summing~\eqref{eq:estimation_error_bound} gives
\begin{align*}
 \sum_{t=0}^{T-1}E_t
 &= E_0\sum_{t=0}^{T-1}\decay^t
       +\sum_{k=0}^{T-2}\omega_k\sum_{t=k+1}^{T-1}\decay^{t-1-k}\notag\\
 &=\Gamma_T(\decay)E_0
       +\sum_{k=0}^{T-2}\Gamma_{T-1-k}(\decay)\omega_k .
\end{align*}
Bounding the geometric sums by $(1-\decay)^{-1}$ gives the stated upper bound. Equation~\eqref{eq:observer_exact} gives the same steps with the actual initial error and the realized increment and innovation norms. Finally, the baseline margin contributes $\bar b\sum_t E_t$ and the robustified margin contributes $S_\sigma\sum_t\omega_t$. Substituting these sums in~\eqref{eq:regret_general} gives~\eqref{eq:design_regret}.

\end{document}